%% file: Draft_RSTA.tex
\documentclass[openacc]{rstransa}%%%%where rstrans is the template name

\jname{rsta}
\Journal{Phil. Trans. R. Soc}
\titlehead{Review article}

\begin{document}

\input{kjtex}

%%%% Article title to be placed here
\title{The profile of the Higgs boson - status and prospects}

\author{%%%% Author details
Karl Jakobs$^{1}$ and Giulia Zanderighi$^{2,3}$}

%%%%%%%%% Insert author address here
\address{$^{1}$Physikalisches Institut, University of Freiburg, Hermann-Herder-Str. 3, 79104 Freiburg, Germany\\ 
https://orcid.org/0000-0001-8885-012X\\
$^{2}$Max-Planck-Institut für Physik, F\"ohringer Ring 6,
80805 M\"unchen, Germany\\% \\ https://orcid.org/0000-0001-6878-1649}
$^{3}$Technische Universit\"at M\"unchen, James-Franck-Strasse 1, 
85748 Garching, Germany\\ https://orcid.org/0000-0001-6878-1649}

%%%% Subject entries to be placed here %%%%
\subject{particle physics / phenomenology}

%%%% Keyword entries to be placed here %%%%

%%%% Insert corresponding author and its email address}
\corres{Karl Jakobs\\
\email{karl.jakobs@uni-freiburg.de}}

%%%% Abstract text to be placed here %%%%%%%%%%%%
\begin{abstract}

The Higgs boson, which was discovered at CERN in 2012, stands out as a
remarkable elementary particle with distinct characteristics. Unlike
any other observed particle, it possesses zero spin within the
Standard Model (SM) of particle physics. Theoretical predictions had
anticipated the existence of this scalar boson, postulating its
interaction with the $W$ and $Z$ bosons as well as through Yukawa
interactions with fermions. Furthermore the Higgs boson can interact
with itself, commonly referred to as the Higgs self-interaction.
In this review, the current state of experimental and theoretical
investigations of Higgs boson production at the LHC and the ongoing
efforts to unravel its properties are described, and an up-to-date
assessment of our understanding of the Higgs sector of the SM is
provided.  In addition, potential links between the Higgs boson and
significant unresolved questions within the realm of particle physics
are presented.
\end{abstract}
%%%%%%%%%%%%%%%%%%%%%%%%%%%

%%%%%%%%%% Insert the texts which can accomdate on firstpage in the tag "fmtext" %%%%%

\begin{fmtext}
\section{Introduction}
The scalar field and its associated particle, now known as the Higgs boson, were postulated in the 1960s~\cite{Englert:1964et,Higgs:1964ia,Higgs:1964pj} 
as a way to introduce massive vector bosons without spoiling the gauge invariance and ensuring the cancellation of unphysical infinities in predictions. 
For nearly fifty years, the realization of this elegantly simple mechanism in nature remained an open question, alongside the unknown value of the Higgs boson mass, denoted as $m_H$.
\end{fmtext}

%%%%%%%%%%%%%%% End of first page %%%%%%%%%%%%%%%%%%%%%

\maketitle

Direct searches for the Higgs boson conducted at the Large Electron Positron collider (LEP) excluded Higgs boson mass values below approximately 114.4 GeV at a 95\% confidence level~\cite{LEPWorkingGroupforHiggsbosonsearches:2003ing}. Global fits of electroweak precision data from LEP and the Tevatron indicated that the Higgs boson should have a relatively light mass, specifically in the range \mbox{115 GeV $\le m_H \le$ 148 GeV}~\cite{Erler:2010wa}.

Soon after the beginning of data taking at the LHC, with data corresponding to an integrated luminosity of only about 5 fb$^{-1}$ at 7 TeV and 6 fb$^{-1}$ at 8 TeV, both the ATLAS~\cite{ATLAS:2012yve} and CMS~\cite{CMS:2012qbp} experiments announced the discovery of a novel scalar particle, the Higgs boson. The discovery primarily involved the production of a Higgs boson through gluon fusion, followed by subsequent decays into either two photons or four charged leptons via two $Z$ bosons. In addition, the decays into two charged leptons and two neutrinos via two $W$ bosons contributed. Early on, several properties of the newfound particle were established, notably its mass, which was observed to be approximately 125 GeV. Moreover, the detection of the Higgs boson's decay into two photons swiftly disfavoured the hypothesis of a new spin-one particle.

The discovery of the Higgs boson stands as an extraordinary achievement in the field of particle physics, both from theoretical and experimental standpoints. The Higgs boson possesses the unique distinction of being the only elementary particle known to possess zero spin, rendering it conceptually and practically distinct from all other known particles. Consequently, extensive efforts from particle theorists and LHC experiments have focused on exploring the Higgs sector of the Standard Model (SM) \cite{Glashow:1961, Weinberg:1967tq, Salam:1968rm}
and accurately determining the properties of the Higgs boson. A noteworthy aspect of the Higgs mechanics within the SM is that once the masses of the Higgs boson and other SM particles are known, all interaction strengths are fully determined. Consequently, any deviation from the SM framework can be interpreted as an indication of New Physics. As a result, precision studies of the Higgs boson and the quest for New Physics are closely intertwined topics.

%=======================================================
\section{The Higgs boson in the Standard Model}

In the SM, the electroweak (EW) symmetry $SU(2)_{L} \times U(1)_{Y}$ undergoes spontaneous symmetry breaking, resulting in the Abelian $U(1)_{\rm em}$ subgroup governing electromagnetism. This process generates the masses of gauge bosons ($W$ and $Z$).

The mechanism responsible for electroweak symmetry breaking (EWSB) in the SM involves a scalar field denoted as $\Phi$, which manifests as an $SU(2)_L$ doublet. The kinetic term and potential of $\Phi$ can be expressed as:
\begin{equation}
{\cal L} \supset | D_\mu \Phi |^2 - V(\Phi)   \label{eq:Lhiggs}\,,
\end{equation}
The first term represents the kinetic term and encompasses the covariant derivative of the $SU(2)_L \times U(1)_Y$ gauge group:
\begin{equation}
D_\mu = \partial_\mu - i g \tau^a W^a_\mu - i g^\prime Y B_\mu\,. 
\end{equation}
Here, $g$ and $g^\prime$ correspond to the $SU(2)_L$ and $U(1)_Y$ gauge couplings, respectively. $\tau^a = \sigma^a/2$, where~$\sigma^a$ are the
usual Pauli matrices and $Y = 1/2$ represents the $U(1)_Y$ hypercharge
of the field $\Phi$.
The physical gauge boson fields ($W^\pm$, $Z$, and $\gamma$) can be related to $W_\mu^a$ and $B_\mu$ through the following equations:
\begin{equation}
\begin{split}
W_\mu^\pm & = \frac{1}{\sqrt{2}} \left (W_\mu^1 \mp i W_\mu^2 \right )\,, \\
Z_\mu & = \cos \theta_w  W_\mu^3 - \sin\theta_w  B_\mu\,,  \\
A_\mu & = \sin \theta_w  W_\mu^3 + \cos \theta_w  B_\mu\,.  
\end{split}
\end{equation}
In these equations, $\sin \theta_w = g^\prime/\sqrt{g^2 + g^{\prime \hspace{0.25mm} 2}} \simeq 0.22$ represents the sine of the weak mixing angle.
The second term in Eq.~\eqref{eq:Lhiggs} is given by the following potential for the scalar field $\Phi$:
\begin{equation}
V(\Phi) = \mu^2 |\Phi|^2 + \lambda |\Phi|^4\,. 
\label{eq:Vhiggs}
\end{equation}
This potential is the most general form that maintains gauge invariance and renormalizability. For $\lambda>0$, when also $\mu^2 \ge 0$, the potential has a unique minimum at $\Phi=0$. However, in nature, $\mu^2<0$, and the minima are located at:
\begin{equation}
|\Phi|^2 = - \frac{\mu^2}{2\lambda} = \frac{v^2}{2}\,. 
\label{eq:Hmin}
\end{equation}
Here, $v = \sqrt{-\mu^2/\lambda}$ represents the vacuum expectation value (VEV) of $\Phi$. Expanding the $\Phi$ field around these minima, in the unitary gauge, it can be expressed as:
\begin{equation}
\Phi = \frac{1}{\sqrt 2} \begin{pmatrix} 0 \\ v + H \end{pmatrix}\,. 
\end{equation}
Here, $H$ is a real scalar, known as the Higgs field. In this gauge, the Higgs potential becomes:
\begin{equation}
V(\Phi) = \frac12 (2 \lambda v^2) H^2 + \lambda v H^3 + \frac{\lambda}{4} H^4\,.
\end{equation}
This implies that the Higgs field $H$ has a mass given by:
\begin{equation}
m_H = \sqrt{2 \lambda} \hspace{0.5mm} v\,. 
\end{equation}
The term involving the covariant derivative of $\Phi$ can also be determined easily, resulting in:
\begin{equation}
|D_\mu \Phi|^2 = \frac12 \hspace{0.25mm} |\partial_\mu H|^2 +\left[\frac{g^2 v^2}{4}  W^+_\mu W^{-\mu} + \frac{\left (g^2 + g^{\prime \hspace{0.25mm} 2} \right ) v^2}{8}  Z_\mu Z^{\mu}\right]\left(1+ \frac{H}{v}\right)^2\,.
\end{equation}
From this expression, the masses of the $W$ and $Z$ bosons can be determined as follows:
\begin{equation}
m_W = \frac{g v}{2} \ ,  \qquad m_Z =
\frac{\sqrt{g^2+g^{\prime \hspace{0.25mm} 2}} \   v}{2}\,. 
\end{equation}

On the other hand, the photon remains massless with $m_A = 0$. The value of the Higgs vacuum expectation value can be determined through precise measurements of the masses of the $W$ and $Z$ bosons, as well as the Fermi constant $G_F \simeq 1.166 \cdot 10^{-5} \  \textrm{GeV}^{-2}$. This is achieved using the relation:
\begin{equation}
\frac{G_F}{\sqrt2} = \frac{g^2}{8 m_W^2} = \frac{1}{2 v^2},
\end{equation}
which is a consequence of matching the Fermi theory to the SM. Numerically, the value of $v$ is found to be approximately $246 \  \textrm{GeV}$. The mass of the Higgs boson is also known with great precision today, as will be discussed in more detail below. By using Eq.~\eqref{eq:Hmin}, which relates the Higgs boson mass and the VEV, one can derive the value of the quartic coupling $\lambda$ in the SM, resulting in $\lambda \simeq 0.13$.
This implies that, in the SM, once the Higgs boson mass is known, the interactions of all vector bosons with the Higgs boson is determined, as well as the Higgs boson self-interactions through triple and quartic Higgs boson couplings.
Similarly in the SM the masses of fermions are generated through the interaction of fermions with the scalar Higgs field and the resulting fermion masses are proportional to the Yukawa couplings of the fermions to the Higgs boson. 
In the following sections, the ongoing theoretical and experimental efforts to test and scrutinize the Higgs mechanism within the SM will be described.

%==============================================================
\section{Theory predictions for Higgs boson production at the LHC}
\label{sec:theory predictions}
Since the discovery of the Higgs boson, the particle physics community has made significant efforts to refine and improve theoretical predictions for Higgs boson production and decay processes at the LHC. Up until 2016, the progress in this field was summarized in four comprehensive "Handbooks." The first handbook focused on inclusive observables~\cite{LHCHiggsCrossSectionWorkingGroup:2011wcg}, the second one addressed differential distributions~\cite{Dittmaier:2012vm}, the third handbook delved into Higgs boson properties~\cite{LHCHiggsCrossSectionWorkingGroup:2013rie}, and the fourth handbook specifically aimed at deciphering the nature of the Higgs sector~\cite{LHCHiggsCrossSectionWorkingGroup:2016ypw}. Since 2016, the field has witnessed significant advancements, and this review primarily highlights the progress made during this period (see also Refs.~\cite{Bass:2021acr,Salam:2022izo}).

A significant part of theoretical advancements concerns the computation of higher-order total cross-sections. These cross-sections are not directly measurable due to inherent fiducial cuts applied in measurements (e.g., due to the limited detector coverage and particle identification criteria). However, these calculations play a crucial role in reweighting fully differential cross sections, which are computed with lower perturbative accuracy, such as cross-sections generated by a parton-shower event simulator. It's worth noting that this reweighting procedure doesn't perform well in specific kinematic regimes and introduces uncertainties that are challenging to quantify. Therefore, the direct computation of fiducial cross sections, as observed by the LHC experiments, is always preferred.

\begin{figure}[htbp]
\begin{center}
     \includegraphics[width=0.85\textwidth]{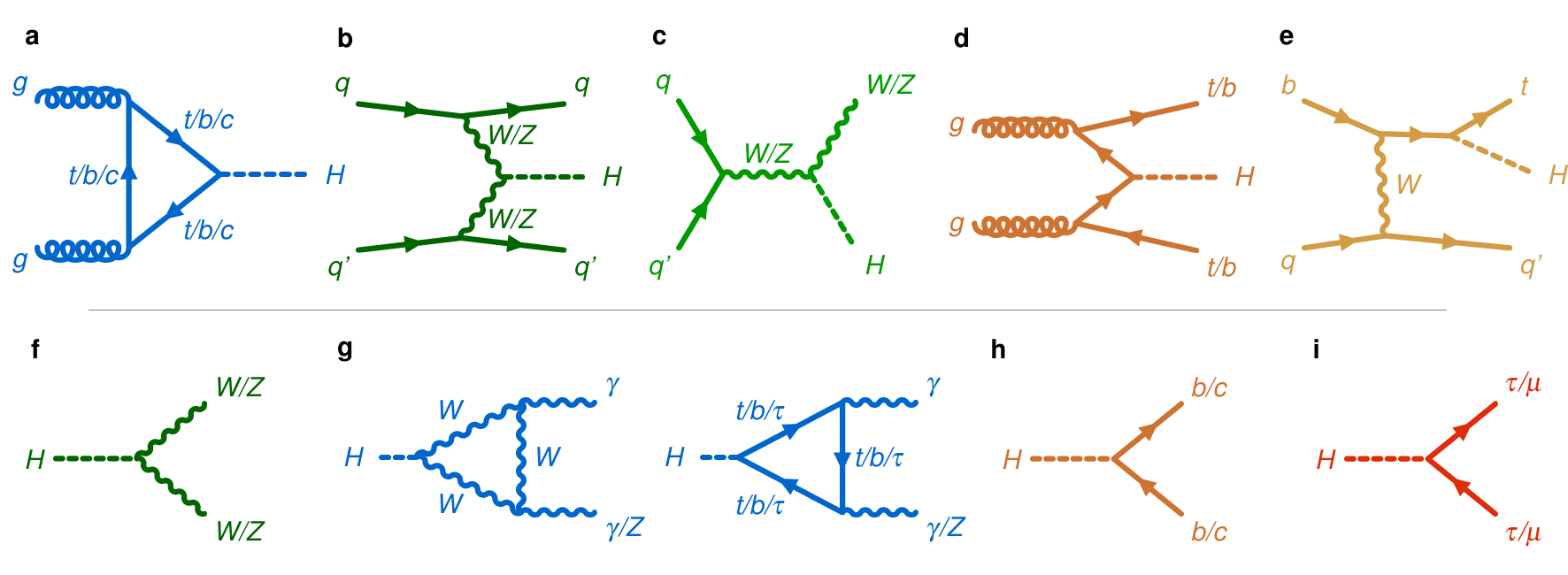}
   \caption{Feynman diagrams for Higgs boson production and decays. a-e: Higgs boson production via (a) gluon fusion, (b) vector boson fusion (c) associated VH production, (d) associated ttH (bbH) production, (e) tH production; \\
   f-i: Higgs boson decays into a pair of vector bosons (f), a pair of photons or a Z boson and a photon (g), a pair of quarks (h) and a pair of charged leptons (i). (Taken from Ref.~\cite{ATLAS:2022vkf}). 
   \label{fig:higgs-prod-dec}}
\end{center}
\end{figure}

\begin{figure}[htbp]
\begin{center}
  \begin{tabular}{lr}
     \includegraphics[width=0.47\textwidth]{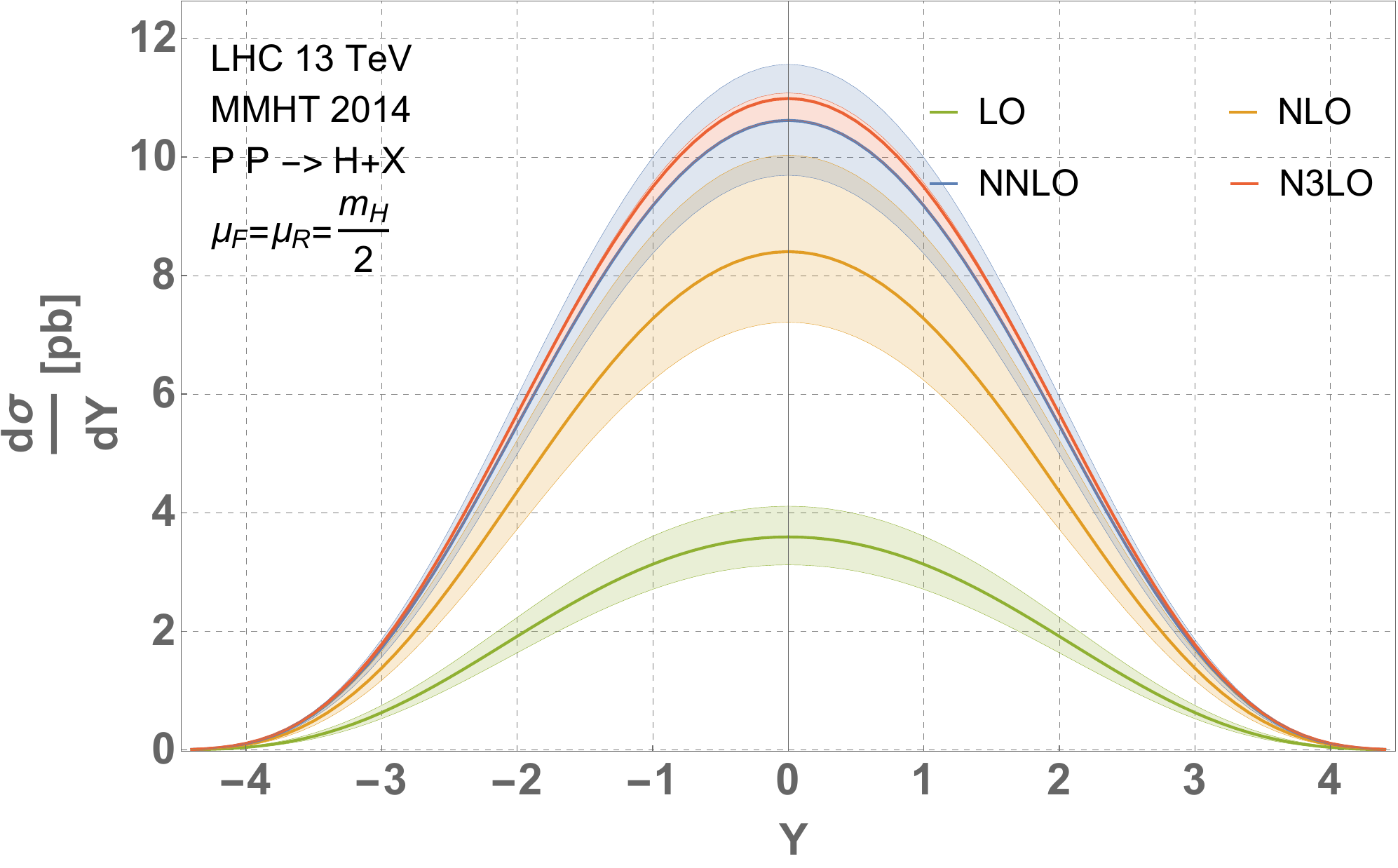}
     &
     \includegraphics[width=0.45\textwidth]{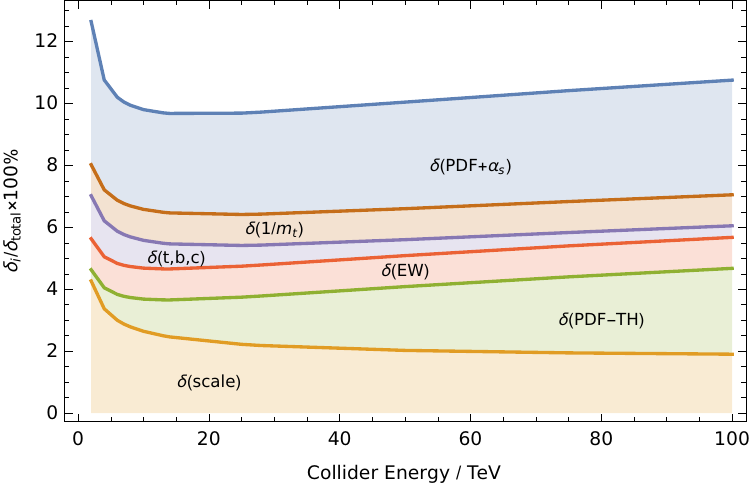}
   \end{tabular}
   \caption{Left: The rapidity distribution of the Higgs boson produced in gluon fusion at different orders in perturbation theory for central renormalization and factorization scales $\mu_R=\mu_F = m_H/2$.  (Taken from Ref.~\cite{Dulat:2018bfe}). Right: Cumulative contributions to the total relative uncertainty on the total Higgs production cross-section, as a function of the collider energy. (Taken from Ref.~\cite{Dulat:2018rbf}). 
   \label{fig:higgs-ggf-theory}}
\end{center}
\end{figure}

\paragraph{Gluon fusion}
The production of the Higgs boson at the LHC involves five major production modes. The main one is known as gluon fusion ($ggF$), where a Higgs boson is produced through the interaction of two incoming gluons mediated by a heavy quark loop (dominated by the top-quark contribution), see Fig.~\ref{fig:higgs-prod-dec}a. This mode has the highest cross-section. The calculation of the inclusive cross-section for this process, within the effective theory where the top quark is integrated out, has been carried out up to three loops in perturbative QCD. Initially, a threshold expansion was performed in Ref.\cite{Anastasiou:2016cez}, and later, exact N$^3$LO effects were included in Ref.\cite{Mistlberger:2018etf}. 
Subsequently, the rapidity distribution was also computed at N$^3$LO in Ref.\cite{Dulat:2018bfe}, and now fully differential N$^3$LO predictions are available in Ref.\cite{Chen:2021isd,Billis:2021ecs}.
Remarkably, this N$^3$LO calculation was the first of its kind for an LHC process. The motivation behind this calculation stemmed from the large corrections observed at NLO and NNLO, as illustrated in Fig.~\ref{fig:higgs-ggf-theory} (left). The N$^3$LO predictions finally stabilize the perturbative expansion for this process, and the N$^3$LO renormalization and factorization uncertainty band, obtained by varying the renormalization and factorization scales independently around the central values $\mu_R=\mu_F=m_H/2$ with the constraint $1/2 \le \mu_R /\mu_F \le 2$, is contained within the NNLO scale variation band.
As far as bottom- and charm-mass effects are concerned, until very
recently, only NLO calculations were
available~\cite{Graudenz:1992pv}. Recently, the interference effects
of top- and bottom-induced production have also been computed at
NNLO~\cite{Czakon:2023kqm}. This calculation is crucial for achieving
precision predictions at the percent level.
Recent research has also estimated the NLO corrections to the mixed QCD-electroweak contribution to Higgs boson production in Ref.\cite{Bonetti:2018ukf}. The study demonstrates that the NLO QCD corrections to the mixed QCD-electroweak contributions are very similar to the NLO QCD corrections for QCD Higgs boson production, validating the factorization approximation of QCD and electroweak corrections. Additionally, the NLO QCD corrections to the light-quark part of the mixed QCD-electroweak contributions to Higgs production via gluon fusion have been recently computed in Ref.\cite{Becchetti:2020wof}.
Despite the high level of sophistication in these calculations, it is crucial to emphasize that the total uncertainty associated with this process is still approximately 10\%, as depicted in Fig.~\ref{fig:higgs-ggf-theory} (right). The figure illustrates the main sources of theoretical uncertainties stacked together as estimated in 2018, as a function of the collider centre-of-mass energy. Since then, progress in the calculation of heavy-quark mass effects~\cite{Czakon:2021yub} has allowed for a reduction in some of these theoretical uncertainties. Nevertheless, the dominant uncertainties at present arise from the parametric uncertainties associated with the strong coupling constant and the parton distribution functions (PDFs).

\paragraph{Vector boson fusion}
The calculation of fully inclusive vector boson fusion (VBF)  Higgs boson production (shown in Fig.~\ref{fig:higgs-prod-dec}b) at N$^3$LO in QCD has been carried out in the approximation 
where the interference between emissions from the two different fermion lines is neglected~\cite{Dreyer:2016oyx}. In this approximation corrections to the VBF process can be computed as corrections to two independent deep inelastic scattering processes, where an incoming fermion scatters off a vector boson exchanged in the $t$-channel. This approximation works well due to the suppression of interference effects by colour and phase space factors at NNLO.

Explicit calculations have confirmed that these non-factorizable corrections, which only arise at NNLO, are indeed at the permille level~\cite{Liu:2019tuy,Dreyer:2020urf,Long:2023mvc,Asteriadis:2023nyl}. In contrast to gluon-fusion Higgs boson production, higher-order corrections for this process are very small. The NNLO corrections to the inclusive cross-section are of the order of a percent, while the N$^3$LO corrections are at the permille level. However, it has been noted in Ref.~\cite{Cacciari:2015jma} within the context of the fully differential NNLO calculation for this process, that corrections in the fiducial phase-space region can be significantly larger.

\begin{figure}[htbp]
\begin{center}
  \begin{tabular}{lr}
     \includegraphics[width=0.47\textwidth]{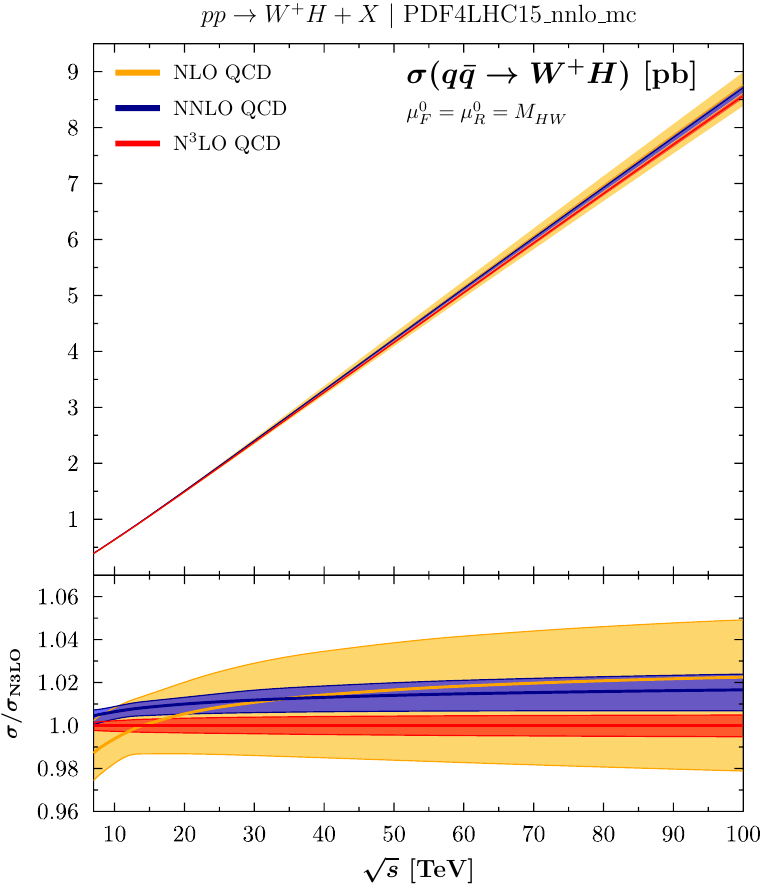}
     &
     \includegraphics[width=0.47\textwidth]{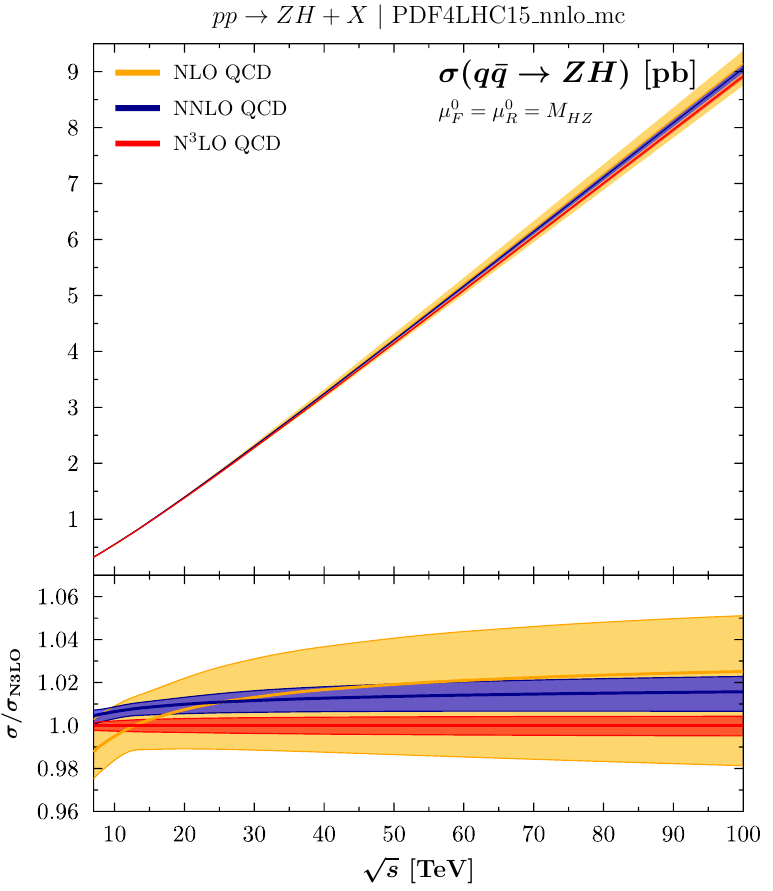}
   \end{tabular}
   \caption{Inclusive cross-section for Higgs boson production in association with a $W^+$ (left) or $Z$ boson (right)
 at a proton-proton collider, as a function of the centre-of-mass energy, up to N$^3$LO in QCD. The uncertainty bands are obtained by performing independent variations of the renormalization and factorization scales around the central values with the constraint $\frac12\le \mu_R / \mu_F \le 2 $.   The lower panels display the ratio to the central N$^3$LO prediction. (Taken from Ref.~\cite{Baglio:2022wzu}). 
   \label{fig:VH-theory}}
\end{center}
\end{figure}

\paragraph{Associated $VH$ production}
Recently, the calculation of the associated production of a Higgs boson with a vector boson 
($V = W, Z$), illustrated in Fig.~\ref{fig:higgs-prod-dec}c, 
at N$^3$LO in perturbation theory has been completed~\cite{Baglio:2022wzu}. Figure~\ref{fig:VH-theory} illustrates the cross-section as a function of the collider energy for a proton-proton collider. The plot clearly demonstrates that the corrections are small, but N$^3$LO effects can extend beyond the uncertainty bands of NNLO theory in this case.

Among the various decay modes in this production channel, one particularly interesting and significant mode is the decay of the Higgs boson into bottom quarks, shown in Fig.~\ref{fig:higgs-prod-dec}h. Thanks to recent advancements, both the production and decay processes can now be described at NNLO, provided that the interference between production and decay is neglected~\cite{Caola:2017xuq,Ferrera:2017zex,Behring:2020uzq}.

A notable distinction between $ZH$ and $WH$ production lies in the fact that in the former case, the final state can be generated at the one-loop level through the fusion of two incoming gluons. Recently, two-loop amplitudes incorporating the full top-quark mass dependence have been computed, and gluon-fusion-induced ZH production is now known up to NLO in QCD~\cite{Hasselhuhn:2016rqt,Chen:2020gae,Davies:2020drs,Wang:2021rxu,Chen:2022rua,Degrassi:2022mro}.

\paragraph{Associated $ttH$  production}

Another major Higgs boson production process involves the production of the Higgs boson in association with two top quarks (see Fig.~\ref{fig:higgs-prod-dec}d). This process is of great importance at the LHC as it allows for a direct measurement of the top quark Yukawa coupling.

While NLO QCD including off-shell effects and EW effects have been known for several years~\cite{Denner:2016wet}, a complete NNLO calculation is currently unavailable due to the lack of knowledge of the full two-loop amplitudes. However, a suggestion was made in Ref.~\cite{Catani:2022mfv} to approximate the onshell cross-section by considering the soft limit of the Higgs boson in the double virtual amplitudes. Despite the fact that this approximation only works moderately well at NLO, since the approximate NNLO correction to the cross-section is very small (of the order of a few percent), even an order one uncertainty on this small correction amounts to an overall small uncertainty on the total cross-section. 

\paragraph{Higgs boson production in association with a single top}

Higgs boson production in association with a single top ($tH$) is the smallest of the major production modes, with a cross-section that is about an order of magnitude smaller than that of $ttH$. Because of this and the signal-to-background conditions, this mode has not been observed yet at the LHC. The importance of this production mode is that, contrary to $ttH$, which provides information only on absolute value of the top Yukawa coupling, it is sensitive to its sign because of the interference between the Higgs boson emission off a $W$ boson and Higgs boson emission off a top quark.
This cross-section is known at NNLO in the five-flavour scheme~\cite{Brucherseifer:2014ama} and at NLO in the four-flavour scheme~\cite{Campbell:2009ss,Demartin:2015uha}. 

\paragraph{Higgs boson pair production}

Double Higgs production at the LHC is a highly significant process as it offers valuable insights into the properties of the Higgs boson. In particular, it allows for the constraint of the Higgs boson self-coupling, which is directly related to the Higgs potential. At the leading order, this process involves a loop of heavy quarks. Triangle diagrams contribute where one Higgs boson couples to the heavy quarks and radiates the second Higgs boson, while box diagrams involve both Higgs bosons emitted from the heavy quarks. The sensitivity of the cross-section to the self-coupling is determined by the triangle diagrams, while the box diagrams reduce this sensitivity.

This process has been computed at N$^3$LO in the heavy-top limit (HTL), where the quark loop is approximated as an effective, point-like interaction~\cite{Chen:2019lzz,Chen:2019fhs}.
Two-loop NLO corrections with the full top-mass dependence have been obtained first in refs.~\cite{Borowka:2016ehy,Buchalla:2018yce}.
Predictions at full NLO QCD in the $\overline{\rm MS}$ scheme for the top-quark mass where also presented in ref.~\cite{Baglio:2018lrj}. This calculation confirmed the results of Ref.~\cite{Borowka:2016ehy}, and showed that a dominant source of uncertainty is due to the top mass and scheme choice. State-of-the-art predictions at NNLO including the top-quark mass and scheme uncertainty can be found in Ref.~\cite{Baglio:2020wgt}. 
Corrections have also been consistently matched to parton showers within the publicly available POWHEG BOX framework~\cite{Heinrich:2017kxx,Heinrich:2019bkc,Heinrich:2020ckp}. The current state-of-the-art description includes approximate NNLO effects in the HTL approximation, enhanced by the insertion of LO form factors accounting for the full top-quark mass dependence, as well as exact NLO corrections in the full theory~\cite{deFlorian:2021azd}. The inclusion of approximate NNLO corrections leads to a reduction in uncertainties by a factor of about 2 to 3.
A fully exclusive Monte Carlo generator within the POWHEG BOX framework which includes consistently NNLO corrections in the heavy top limit is also available~\cite{Alioli:2022dkj}.

\section{Higgs boson production and decays at the LHC}

\subsubsection{Higgs boson discovery and decays into bosons}

The discovery of the Higgs boson was announced in 2012 by the experiments ATLAS \cite{ATLAS:2012yve} and CMS \cite{CMS:2012qbp} at the Large Hadron Collider (LHC) \cite{Evans:2008zzb} only two and a half years after the beginning of data taking. It was based on the analysis of proton-proton ($pp$) collision data collected during the years 2011 and 2012 at centre-of-mass energies of 7~TeV and 8~TeV, respectively. The integrated luminosity, which is directly proportional to the number of produced Higgs bosons, was about 5~$\fbs$ at 7 TeV and about 6~$\fbs$ at 8 TeV in the two experiments. 

The discovery was largely based on the search for Higgs boson decays into bosons leading to photons and leptons in the final state, i.e. $\hgg$, $\hzzssfourl$,  $\hwwssll$ (see Fig.~\ref{fig:higgs-prod-dec}~f,g). Although the combined branching ratios for these decays are small, their choice was driven by the expected signal-to-background levels. In addition, the $\hgg$, $\hzzssfourl$ decays allow to reconstruct the Higgs boson mass as the invariant mass of the final state photons or leptons and the Higgs boson appears as a sharp resonance -- the width of which is dominated by the detector resolution -- on top of smooth backgrounds, which are dominated by \gamgam\ and $ZZ^{*}$ continuum production, respectively. 

Since the announcement of the discovery the available $pp$ collsion data have been significantly increased during the second data-taking period (Run 2) between 2015 and 2018. In total, data corresponding to an integrated luminosity of about 140~$\fbs$ could be recorded by each of the experiments at the increased centre-of-mass energy of 13~TeV. The collection of this large dataset is due to the sensational performance of the LHC, well beyond expectations, as well as due to the excellent performance of the experiments in terms of data-taking efficiency and data quality. Overall, the combination of the performance of the LHC machine, the detectors and the GRID computing have proven to be a terrific success story. 

\begin{figure}[htbp]
\begin{center}
%     \includegraphics[width=0.95\textwidth]{figures/Fig9_links_new_b}
%  \begin{tabular}{lr}
     \raisebox{2.125ex}{\includegraphics[width=0.45\textwidth]{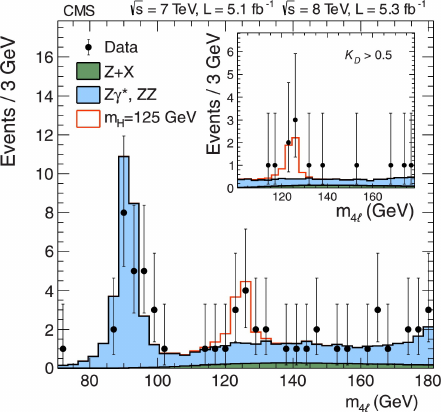}} \qquad
%     &
     \includegraphics[width=0.48\textwidth]{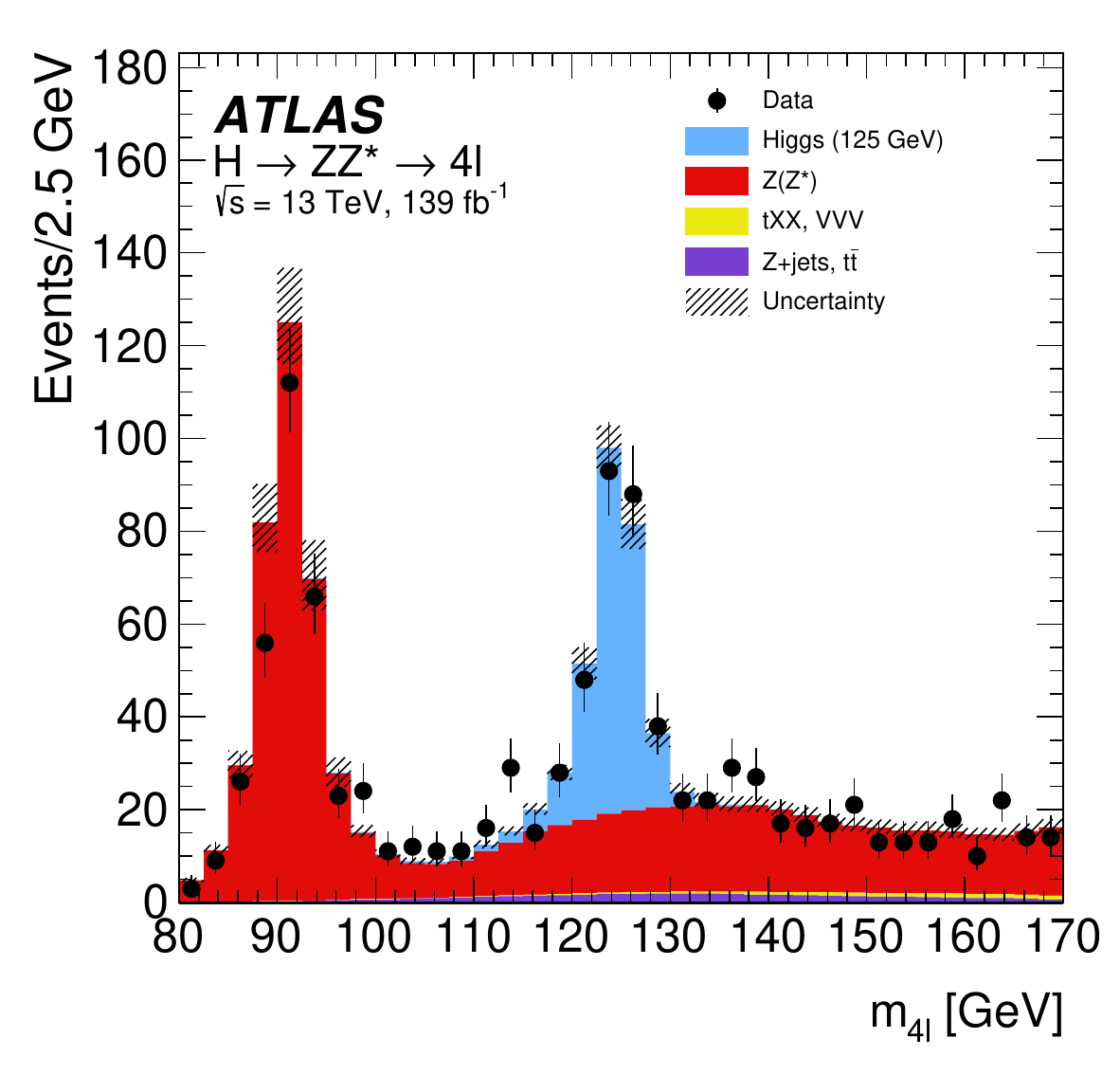}
%   \end
   \caption{Distributions of the invariant mass of selected events with two
    pairs of unlike-sign electrons or muons measured by the CMS experiment at the time of the Higgs boson discovery \cite{CMS:2012qbp} (left) and
    by the ATLAS experiment using the full Run-2 data \cite{ATLAS:2020wny} (right).
   \label{fig:higgs-4l-run1-run2}}
\end{center}
\end{figure}

\begin{figure}[htbp]
\begin{center}
  %\begin{tabular}{lr}
     \raisebox{0.9ex}{\includegraphics[width=0.48\textwidth]{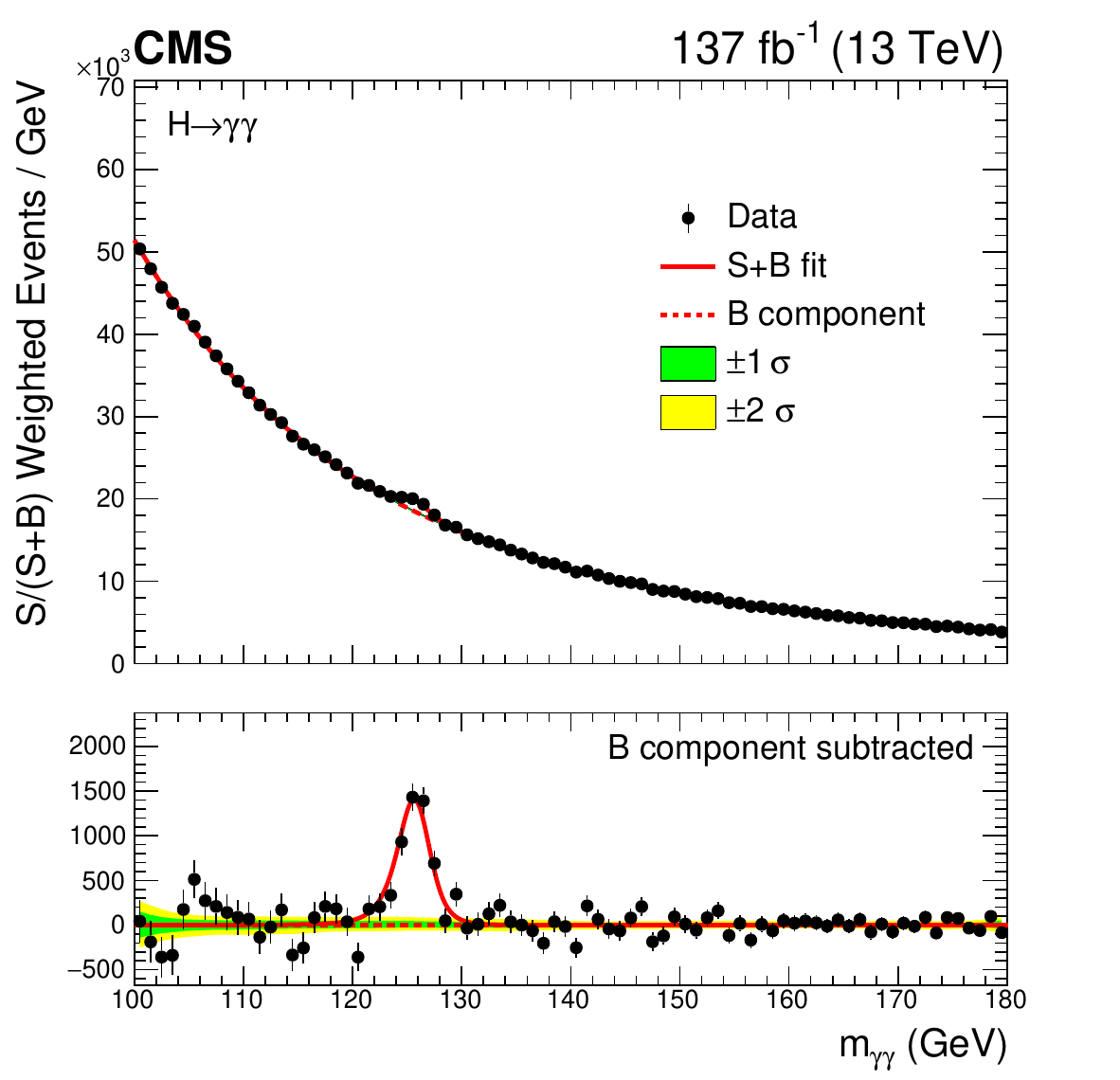}}\qquad
   %  &
     \includegraphics[width=0.35\textwidth]{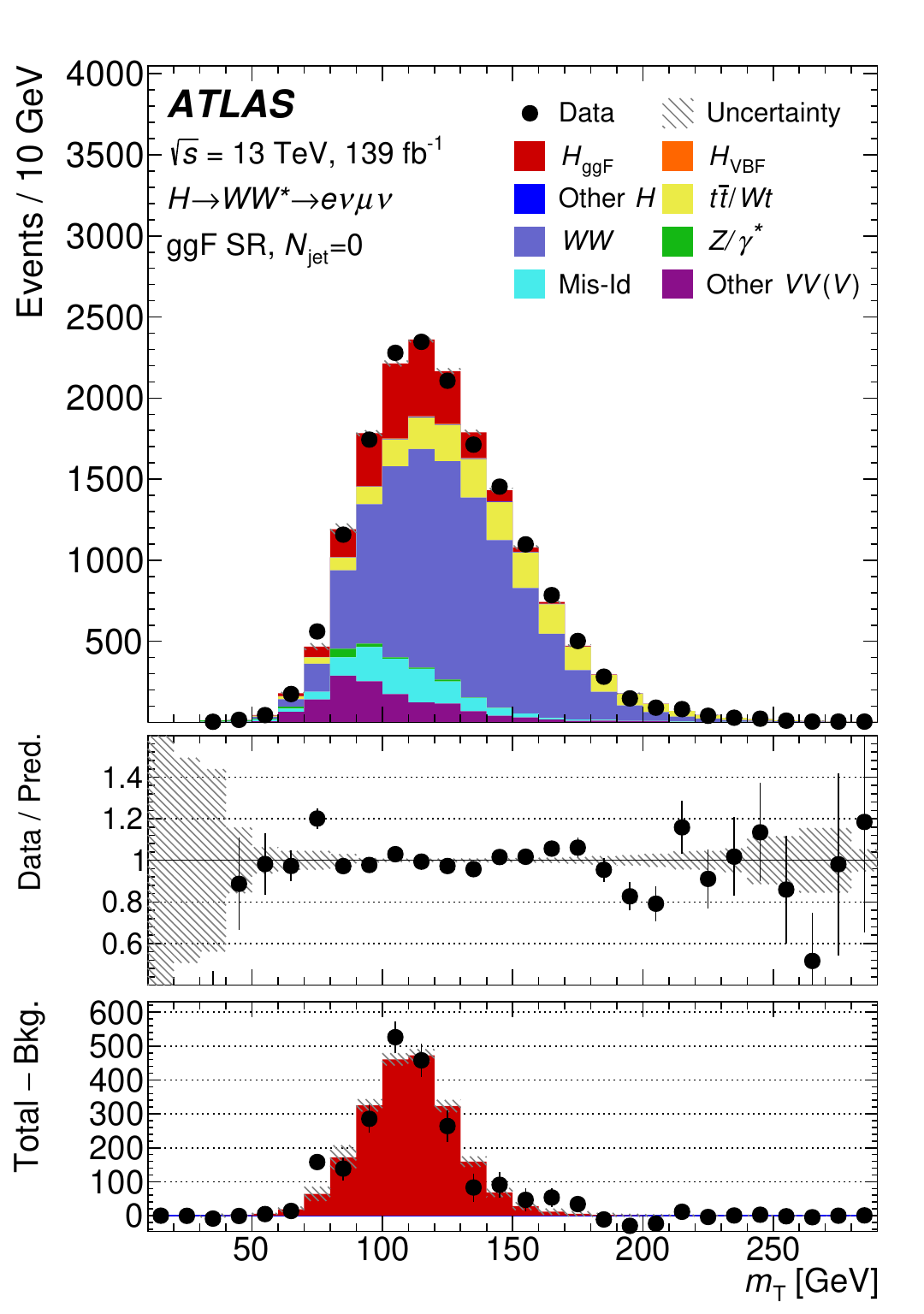}
   %\end{tabular}
   \caption{Distributions of the invariant mass of selected Run-2 events with two photons 
    measured by the CMS experiment \cite{CMS:2022wpo} (left) and of the transverse mass of events with one electron, one muon and missing transverse momentum by the ATLAS experiment \cite{ATLAS:2022ooq} (right).}
   \label{fig:higgs-bosons-run2}
\end{center}
\end{figure}

The higher rate of collisions as well as the increased cross-sections for Higgs boson production at 13~TeV led to a large sample of recorded Higgs boson decays, 30 times larger than at the time of its discovery, which represents a unique sample to study the properties of this particle in greater detail and with increased precision. The amazing difference at the level of the reconstructed Higgs boson signal can be seen in Fig.~\ref{fig:higgs-4l-run1-run2}, where distributions of the invariant mass of the selected four-lepton final states are shown at the time of discovery and for the complete Run-2 data. While at times of the discovery the three channels mentioned above had to be combined to reach a statistical significance of 5$\sigma$, clear signals are now established in each of these bosonic decay modes. In Figure~\ref{fig:higgs-bosons-run2} the reconstructed di-photon invariant mass \cite{CMS:2022wpo} and the reconstructed transverse mass \cite{ATLAS:2022ooq} are shown for selected events in the $\hgg$ and  $\hwwssll$ decays modes for the CMS and ATLAS experiments, respectively. 
The transverse mass is defined as 
$m_T = \sqrt{2 \ p_T^{\ell \ell} \ E_T^{\rm{miss}} \ ( 1 - \cos{\Delta \phi}) }$, where $p_T^{\ell \ell} $ represents the transverse momentum of the di-lepton system, $E_T^{\rm{miss}}$ is the missing transverse energy, and $\Delta \phi$ is the azimuthal separation between the directions of the di-lepton system and the missing transverse energy.  
In each of these channels, the signal significance exceeds clearly the 5$\sigma$ threshold. By exploiting additional signatures in the final states, such as forward jets for VBF production or additional leptons for $WH$ and $ZH$ production the different production processes can also be separated. Both experiments have meanwhile established signals for the four major production modes ($ggF$, VBF, $WH$ and $ZH$, and $\ttbar H$ production) via decays in bosonic and fermionic (see below) channels. The separation in different production and decay modes plays an important role in the determination of Higgs boson couplings to the SM particles, as discussed below. 

\subsubsection{Higgs boson decays into fermions}

In order to establish the mass generation for fermions as implemented in the SM, it is of prime 
importance to demonstrate the direct couplings of the Higgs boson to fermions and their proportionality to mass. The most prominent candidate decay modes are the decays into $\tau$ leptons, $\htautau$  (Fig.~\ref{fig:higgs-prod-dec}i), and into $b$ quarks, $\hbb$ 
(Fig.~\ref{fig:higgs-prod-dec}h)
Given the overwhelming background from multi-jet production via QCD processes, the $\bbbar$ channel requires the restriction to Higgs bosons produced in association with vector bosons or $\ttbar$ pairs. In these cases, either the leptonic decay of 
the vector boson or of the top quark(s) may provide high-$\pt$ leptons, such that the events can be triggered and the signal-to-background ratios improved. More favourable signal-to-background conditions exist for \htautau\ decays.

After Run 1, evidence for fermionic decays in the \htautau\ channel was presented with significances of 3.2$\sigma$ \cite{CMS:2014wdm} and 4.5$\sigma$  \cite{ATLAS:2015xst} in the CMS and ATLAS experiments, respectively. In a combination of the results of the two experiments a significance of 5.5$\sigma$ was obtained \cite{ATLAS:2016neq}. Independent observation in each of the two experiments of the direct coupling of the Higgs boson to fermions was observed during Run 2. The observation of the couplings to the quarks and charged leptons of the third generation ($t$- and $b$-quarks and $\tau$-leptons) is one of the main achievement of the LHC experiments in Run 2.  

\begin{figure}[htbp]
\begin{center}
  %\begin{tabular}{lr}
     \raisebox{1ex}{\includegraphics[width=0.45\textwidth]{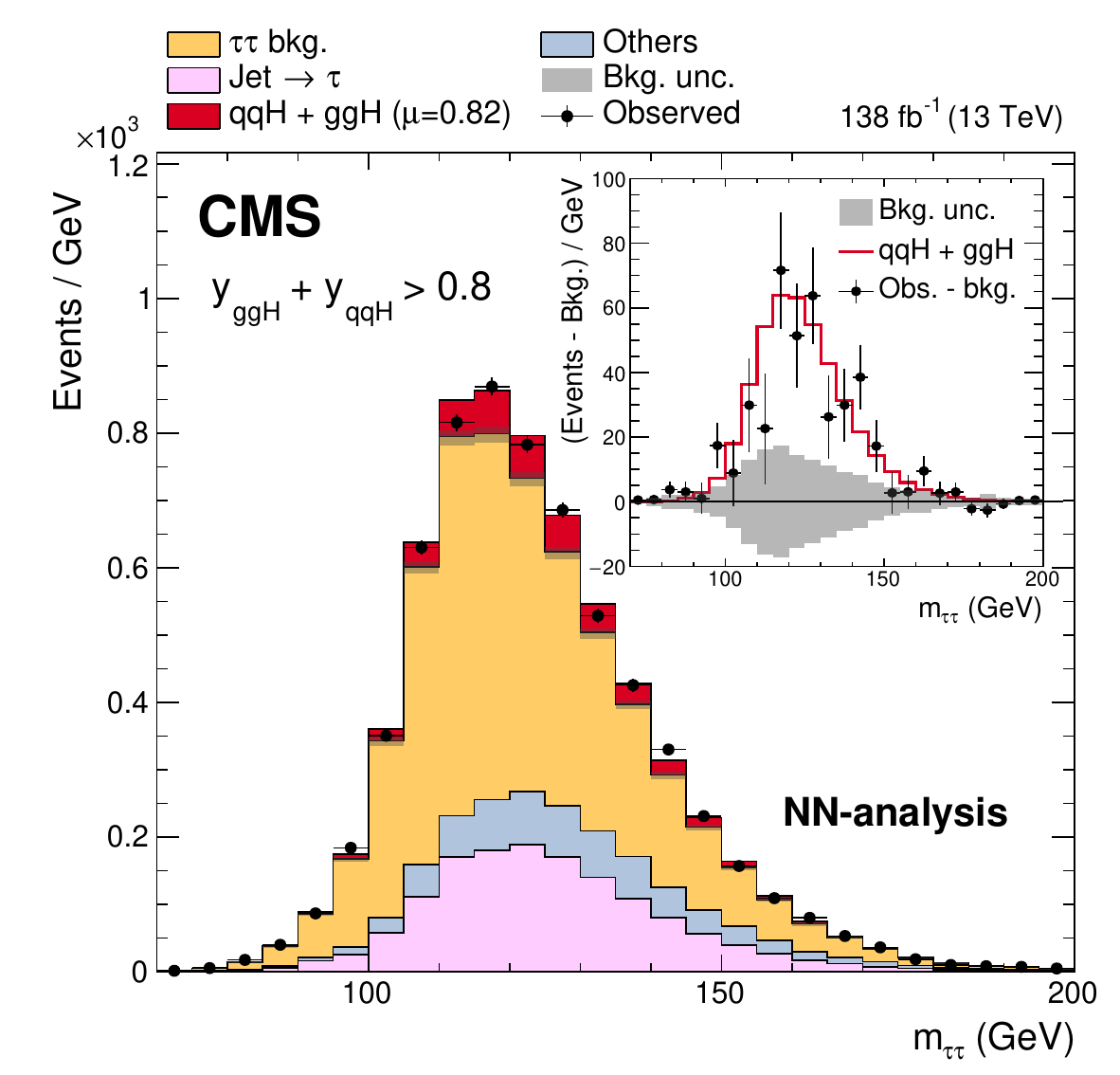}}\qquad\qquad
     %&
     \includegraphics[width=0.45\textwidth]{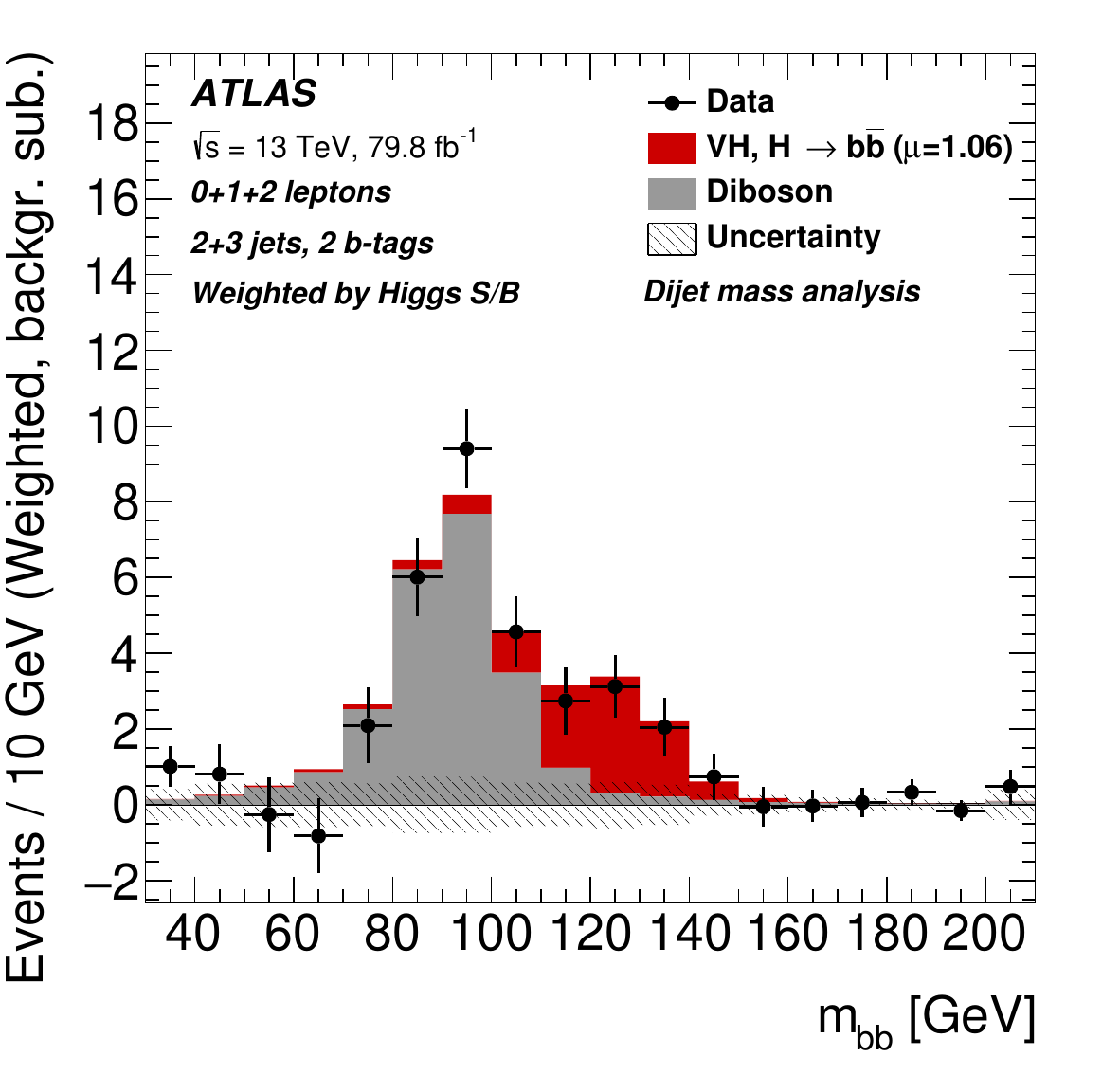}
   %\end{tabular}
   \caption{Distributions of the invariant mass of selected Run-2 events with two tau leptons
    measured by the CMS experiment \cite{CMS:2022kdi} (left) and of the invariant mass of two b quarks, $m_{bb}$ by the ATLAS experiment \cite{ATLAS:2018kot} (right).}
   \label{fig:Run2-Hfermions}
\end{center}
\end{figure}

In the search for $\htautau$ decays all combinations of leptonic ($\tau \to \ell \nu   \bar  \nu$, with $\ell = e, \mu$)
and hadronic ($\tau \to \text{hadrons} \ \nu$) $\tau$  decays were considered.
The search was designed to be sensitive to the major production processes of a SM Higgs boson.
The different production processes led to different final-state signatures, which have been
exploited by defining an event categorisation.
Dedicated categories (VBF and high-\pt\ categories) were considered to achieve both a good signal-to-background ratio and a good resolution for the reconstruction of the $\tau \tau$ invariant mass. Given the challenging signal-to-background conditions, and in order to exploit correlations between final-state observables, multivariate analysis techniques
based on Boosted Decision Trees were used. Important backgrounds were normalised in dedicated control regions and transported to the respective signal regions. Both collaborations published the observation of this decay mode already using partial Run-2 datasets \cite{CMS:2017zyp , ATLAS:2018ynr}. As an example, the reconstructed invariant mass of the two $\tau$  leptons, \mtautau, for the most recent CMS analysis \cite{CMS:2022kdi}, is shown in Fig.~\ref{fig:Run2-Hfermions}. 

The LHC physics highlights of the year 2018 were the observations of the couplings of the Higgs boson to the quarks of the third generation. In the same year the ATLAS and CMS collaborations observed the associated $\ttbar H$ production 
\cite{CMS:2018uxb , ATLAS:2018mme}, which provides direct access to the top-Higgs Yukawa coupling, and the Higgs boson decays into a pair of b-quarks, \hbb\ \cite{ATLAS:2018kot, CMS:2018nsn}. Also these results were achieved due to the excellent performance of the experiments and the application of sophisticated analyses methods based on multivariate techniques. Before the startup of the LHC the observation of these channels was considered to be extremely challenging and was estimated to require huge datasets and a careful control of systematic uncertainties on the large backgrounds. 

The $\hbb$ decay mode is predicted in the SM to have a branching ratio of 58\% for \mH\ =125~GeV. 
Accessing $\hbb$ decays is therefore crucial for constraining the overall Higgs boson decay width. Due to large backgrounds from multi-jet production, a sufficiently large signal can only be observed in the associated production with a vector boson ($VH$ production). The analyses were performed for events containing zero, one, or 
two charged leptons (electrons or muons), targeting the $Z \to \nu \nu$, $W \to \ell \nu$ or $Z \to \ell \ell$ decay 
modes of the vector boson, respectively. Highly performing $b$-tagging algorithms were used to identify jets consistent with the hypothesis of them originating from the hadronisation of a 
$b$-quark. 
To improve the sensitivity, the three channels
are each split according to the vector-boson transverse momentum, the
number of jets (two or three), and the number of $b$-tagged jets. Topological and kinematic selection criteria are applied
within each of the resulting categories.

Based on data collected until the end of 2017, corresponding to an integrated luminosity of about 80~\fbs\ and in combination with the results based on the Run-1 data, the \hbb\ decay mode was observed with significance of 5.4$\sigma$ and 5.6$\sigma$ by the ATLAS \cite{ATLAS:2018kot} and CMS \cite{CMS:2018nsn} collaborations, respectively. Additionally, a combination of Run-2 results searching for the Higgs boson produced in association with a vector boson established at the same time the observation of the $VH$ production mode \cite{ATLAS:2018kot}.

Even more challenging is the search for couplings of the Higgs boson to fermions of the second generation. In the complete Run-2 dataset only about 2000 \hmumu\ events (Fig.~\ref{fig:higgs-prod-dec}i) are expected to be produced on top of a huge background from Drell-Yan production. Despite this, both collaborations have seen interesting indications for the presence of such decays, with observed significances of 3.0$\sigma$ \cite{CMS:2020xwi}  and 2.0$\sigma$ \cite{ATLAS:2020fzp} in the CMS and ATLAS experiments, respectively. To reach these sensitivities, categorisation into different topologies and the use of BDT discriminants were necessary. The observed rates are found to be in agreement with the expectations for a SM Higgs boson. 

The collaborations have also started to address the next challenge, the search for $\hcc$ decays (Fig.~\ref{fig:higgs-prod-dec}h). Like the observation of the \hbb\ decay mode, this search is based on the associated production of the Higgs boson with a vector boson ($VH$ production), such that leptons provide both the trigger signature and a significant background rejection. A key asset in this search is a strong identification of $c$-quark jets and their separation from $b$-quark and light-quark and gluon jets. Novel charm taggers and analysis methods have been developed based on machine-learning techniques. Given the analysis challenges and the small couplings to the relatively light c-quark, the Run-2 data do not yet provide sensitivity to reach the expected SM rates. However, limits on the observed signal strength of $\hcc$ contributions have been set  
by the CMS and ATLAS collaborations to correspond to 14 \cite{CMS:2022psv} and 26 \cite{ATLAS:2022ers} times the SM values, respectively.  

In summary, Higgs boson couplings to fermions of the third generation have been clearly established in Run 2 at the LHC and first steps towards a measurements of couplings to second-generation fermions have been made and the results look very promising. More data from Run 3 and beyond are essential to firmly establish these couplings and to measure their values. 

\subsubsection{Rare and invisible Higgs boson decays}
In addition to the decay modes discussed above, searches for the decays $\hZg$ (Fig.~\ref{fig:higgs-prod-dec}g), with $Z \to \ell \ell$ were performed. This channel has a very low branching ratio in the SM, but new physics might lead to a significant enhancement, 
such that it might become visible. In the analyses of the ATLAS and CMS collaborations first indications of a small excess have been seen, corresponding to observed significances of 2.2$\sigma$ \cite{ATLAS:2020qcv} (ATLAS) and 2.7$\sigma$ \cite{CMS:2022ahq} (CMS). Within the large uncertainties the observed excesses above the backgrounds are in agreement with the expectations for SM \hZg\ decays.  By combining the results of the two collaborations, first evidence for this decay mode with a significance of 3.4$\sigma$ 
\cite{CMS:2023mku} has been obtained. 

Important knowledge on the profile of the discovered Higgs boson is contributed by searches for invisible final states. Within the SM, decays to neutrinos via the $ H \to Z Z^*$ channel constitute a well-known albeit very small contribution. 
However, searches for invisible decays are also sensitive to additional decay
modes into new particles, which are predicted in some theories beyond the SM. 
Relatively clean signatures are provided by the VBF and $ZH$ production channels, where the accompanying tag jets and leptons from the $Z$ decay provide the final state signatures together with large missing transverse momentum resulting from the invisible Higgs boson decay. Since no excess above the expected backgrounds from SM processes are observed, 95\% CL upper limits on the cross section times branching ratio for a Higgs boson decaying into invisible particles have been set.  
Assuming the SM production cross section for a Higgs boson with a mass of 125~GeV, these limits have been translated into constraints on the branching fraction to invisible particles of the discovered Higgs boson. The most stringent bounds on the invisible branching fraction of the Higgs boson have been extracted from a combination of searches in various channels and have been found to be 10.7\% in the ATLAS experiment \cite{ATLAS:2023tkt} and 18\% in the CMS experiment \cite{CMS:2022qva}.

\section{Higgs boson properties}

\subsubsection{Mass}
The Higgs-boson mass was the only remaining unknown parameter in the SM prior to the discovery.
The LHC collaborations have chosen a model-independent approach to measure the Higgs boson mass
based on fitting the spectra of the reconstructed invariant masses of the \hgg\ and
\hzzssfourl\ decay modes. In these two channels the Higgs boson shows up as a narrow mass peak with a typical experimental resolution of 1.6 to 2~GeV over a smooth background.
The results on the Higgs boson mass depend on the precise calibrations 
of the energy scale of the electromagnetic calorimeter and on the muon momentum scale. In the \hgg\ channel, the mass measurement requires good control of the energy scale and resolution of photons, which are coupled to the electron 
calibration.  

\begin{figure}[htbp]
\begin{center}
  %\begin{tabular}{lr}
     \includegraphics[width=0.40\textwidth]{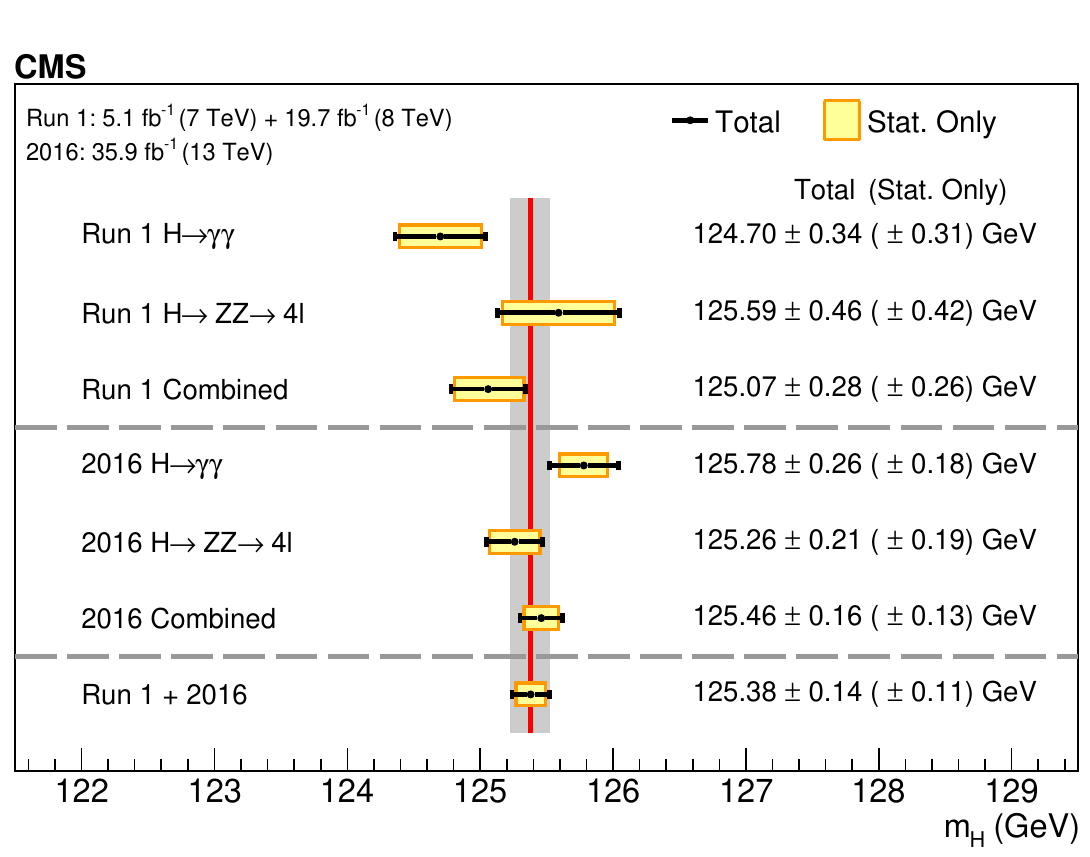}
\qquad 
     %&
     \includegraphics[width=0.47\textwidth]{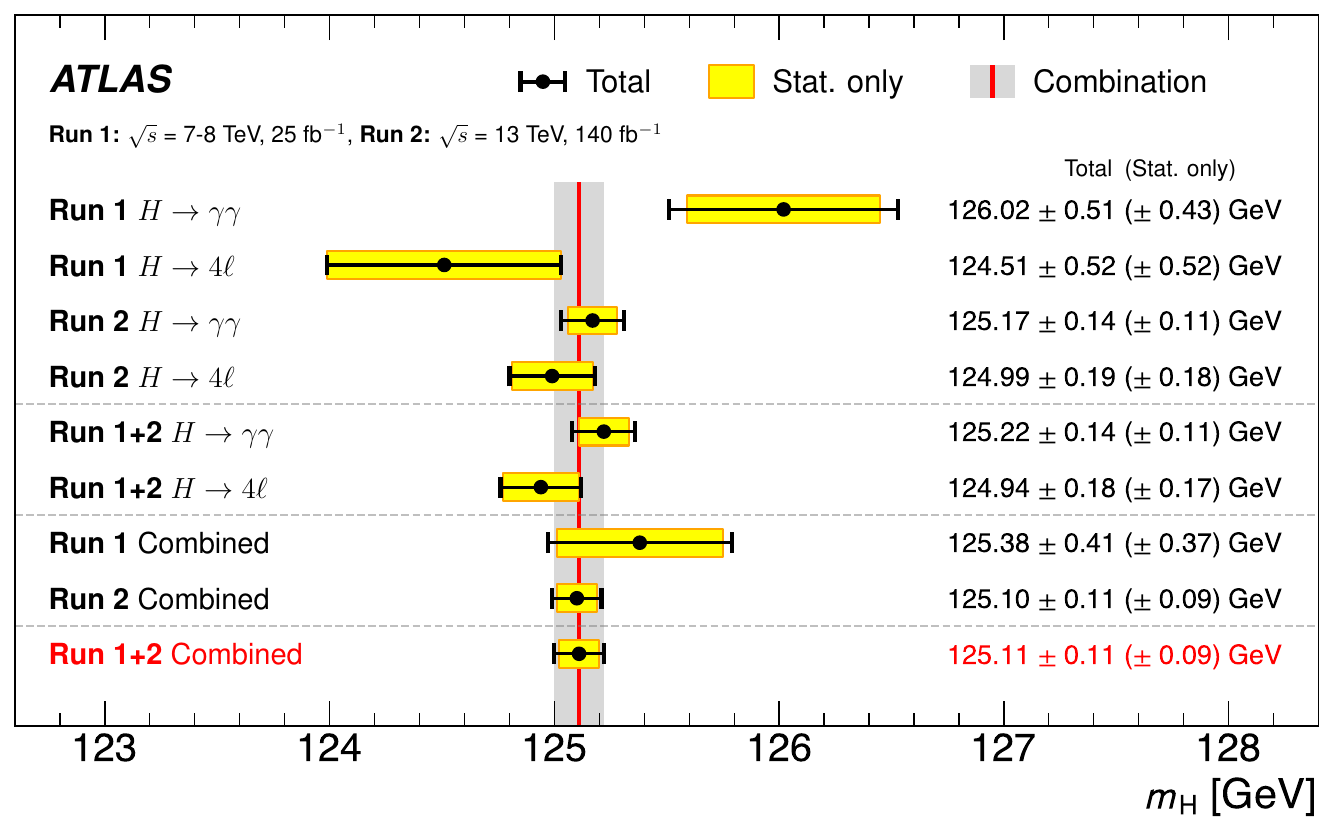}
   %\end{tabular}
   \caption{Results on the Higgs boson mass measurements in various final states for the CMS \cite{CMS:2020xrn} (left) and the ATLAS experiment \cite{ATLAS:2023oaq} (right). The uncertainty bar on each point is the total statistical and systematic uncertainty; the numbers in the brackets show the statistical uncertainty only. The vertical (red) lines indicate the final combined results, and the grey bands their total uncertainties.}
   \label{fig:Higgs_mass}
\end{center}
\end{figure}

An overview of different Run-1 and Run-2 measurements as well as the result of the combination by the CMS experiment \cite{CMS:2020xrn} is shown in Fig.~{\ref{fig:Higgs_mass}} (left). The results of the ATLAS measurements \cite{ATLAS:2023oaq} are displayed in the right part of Fig.~{\ref{fig:Higgs_mass}}. 
The combined Run-1 and Run-2 result from the ATLAS experiments includes already the full Run-2 dataset for both the \hzzssfourl\ \cite{ATLAS:2022net} and \hgg\ \cite{ATLAS:2023owm} decay modes. It yields a Higgs boson mass of 125.11 $\pm$ 0.11 GeV, which corresponds to a precision of 0.09\% on this fundamental parameter of the SM. The precision in the four-lepton channel is still dominated by the statistical uncertainty, which implies that on the longer term this channel will become the dominant one to obtain the final precision. 

\subsubsection{Spin and CP properties}
In the SM the Higgs boson is a CP-even spin-0 state ($s^{CP}=0^{+}$),  where CP denotes the combined operation of parity (P) and charge conjugation (C). In theories beyond the SM, also CP-odd states can
occur in the Higgs sector, and in general there is also the possibility 
that a state is not a CP eigenstate but rather a mixture of CP-even and CP-odd components.
Experimentally, under the assumption that the observed signal corresponds to a single resonance rather than an overlap of several states with different spins, the spin of the observed particle have been probed by testing distinct hypotheses for spin 0, 1 and 2. 

Information on spin and CP properties have been obtained in particular from the analysis of angular relations between either initial-state or final-state objects. The $\hzzssfourl$ channel is the most powerful channel and offers excellent
sensitivity to the spin of the decaying particle by exploiting the angular distributions of the lepton pairs. All results obtained have been found to be consistent with the expectations for $s^{CP}=0^{+}$ and a large number of alternative spin-parity hypotheses have been ruled out with a confidence level exceeding 99\% \cite{CMS:2014nkk, ATLAS:2015zhl}. 

\subsubsection{Higgs boson width}
Within the SM, the total width of the Higgs boson with a mass of 125 GeV is expected to be about 4.1~MeV, which is several orders of magnitude below the experimental resolution even in the high-resolution $\hgg$ and $\hzzssfourl$ channels. 

To extract information on the Higgs boson width $\Gamma_H$ in an indirect way, a method relying on measurements of both off-shell and on-shell Higgs boson production has been developed
\cite{Kauer:2012hd, Caola:2013yja}. While on-shell Higgs boson production is inversely proportional to the width, the off-shell production is largely independent of it: 

\begin{equation}
\nonumber
\sigma^{\rm on-shell}  \sim  \frac{g^2_p g^2_d}{\Gamma_H}\,,   \hspace{2.0cm}
\sigma^{\rm off-shell} \sim   {g^2_p g^2_d} \ ,
\label{eq:Higgs_width}
\end{equation}
where $g_p$ and $g_d$ are the couplings associated with the Higgs boson production, which is dominated by gluon fusion, and decay.
Therefore, the measurement of the ratio of the two cross-sections provides information on $\Gamma_H$. However, it has to be assumed that the coupling constants in the on-shell and off-shell regimes are given by the SM and that no new particles enter into the gluon-fusion process. 

Several searches for off-shell Higgs boson production have been carried out by the ATLAS and CMS collaborations using Run-1 and Run-2 data, where mainly $H \to ZZ$ decays have been considered. The measured off-shell cross-section is extracted by taking into account the negative interference of the continuum background from $gg \to ZZ$ 
and the off-shell Higgs boson production.

In recent publications, the CMS and ATLAS collaborations have reported evidence for such off-shell contributions to the production cross-section of two Z bosons. The width of the Higgs boson was determined to be $\Gamma_H$ = 3.2 $\pm \ $2.4~MeV
\cite{CMS:2022ley} and $\Gamma_H = 4.5 ^{+3.3}_{-2.5}$~MeV 
\cite{ ATLAS:2023dnm} by the CMS and ATLAS collaborations, respectively. These determinations are in excellent agreement with the SM expectation and -- albeit indirect -- provide important information on the lifetime of the Higgs boson.

\subsubsection{Higgs boson couplings to SM particles}

The Higgs boson production rates are probed by a likelihood fit to observed signal yields in the various production and decay modes. Because the production cross-section $\sigma_i$ and the branching fraction $B_f$ for a specific production process $i$ and decay mode $f$ cannot be measured separately without further assumptions, the observed signal yields are expressed in terms of signal strength modifiers
$\mu_{if} = (\sigma_i / \sigma_i^{SM} ) (B_f / B_f^{SM})$, where the superscript 'SM' denotes the corresponding SM prediction. 

\begin{figure}[htbp]
\begin{center}
    \includegraphics[width=0.95\textwidth]{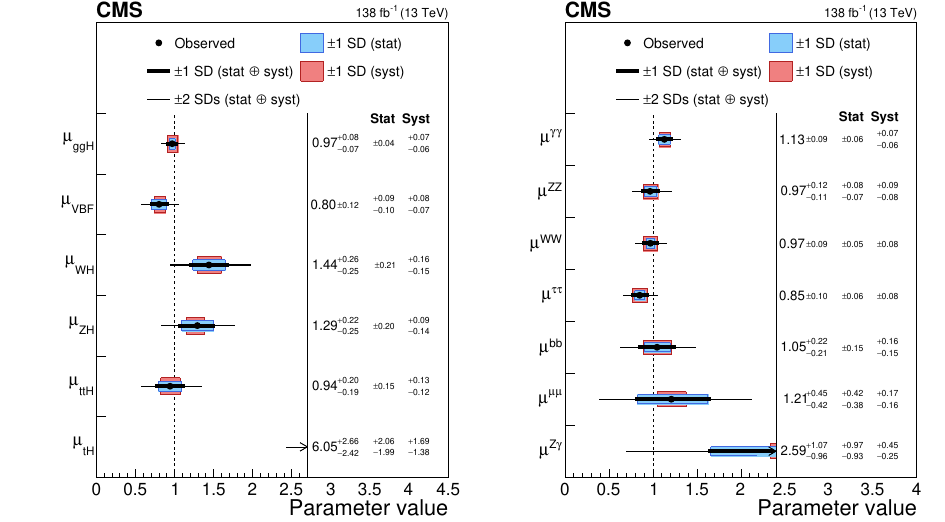}
   \caption{Ratios between the observed and predicted Higgs boson production cross-sections and branching fractions, as measured by the CMS collaboration \cite{CMS:2022dwd}. Left: ratios of the cross sections for different production processes are measured assuming SM values for the decay branching fractions. Right: ratios of branching fractions for different Higgs boson decay modes are measured assuming SM values for the production cross sections. 
   \label{fig:combination_mu_prod-decay}}
\end{center}
\end{figure}

A first test of compatibility is performed by fitting the $\mu$ values for the different production processes and decay modes. The results, as determined by the CMS collaboration 
\cite{CMS:2022dwd}, are presented in Fig.~\ref{fig:combination_mu_prod-decay}. The production cross sections are measured assuming SM values for the decay branching fractions and vice versa, i.e. assuming SM values for the production cross-sections, when determining the ratios of branching fractions. All measurements are compatible with the SM predictions and all major production processes have been established with significances exceeding 5$\sigma$.  

To determine the value of a particular Higgs boson coupling strength, a simultaneous fit of many individual production cross-section times branching fraction measurements is required. The coupling fit presented here is performed within the $\kappa$ framework \cite{LHCHiggsCrossSectionWorkingGroup:2013rie} with a set of multiplicative coupling strength modifiers $\kappa$ that affect the Higgs boson coupling strengths without altering any kinematic distributions of a given process.
The coupling scale factors $\kappa_j$ are defined in such a way that the cross sections
$\sigma_{j}$ and the partial decay widths $\Gamma_{j}$ associated with the SM particle $j$ scale
with $\kappa_j^2$ compared to the  SM 
prediction~\cite{LHCHiggsCrossSectionWorkingGroup:2013rie}.
With this notation,
 and with $\kappa_{H}^2$ being the scale factor
 for the total Higgs boson width $\Gamma_{H}$,
 the cross section for the $gg \to H \to \gamgam$ process, for example, can be expressed as:
\begin{eqnarray*}
  \frac{\sigma\cdot \text{B}~( gg \to H \to \gamgam)}{\sigma^\text{SM}(gg \to H) \cdot \text{B}^\text{SM}(H \to \gamgam)}
  &=&
  \frac{\kappa_{g}^2 \cdot \kappa_{\gamma}^2}{\kappa_{H}^2}\label{eq:cross} .
\end{eqnarray*}

By definition, the currently best available SM predictions for all
$\sigma\cdot \text{B}$, including higher-order QCD and electroweak
corrections, are recovered when all $\kappa_j = 1$. The parametrisation takes into account that the total decay width of the Higgs boson depends on all decay modes included in the present measurements, as well as currently undetected or invisible decays predicted by the SM (such as those to gluons, light quarks or neutrinos) and the hypothetical decays into non-SM particles. 
The corresponding branching fractions for the two are denoted by $B_{\rm{inv}}$ and $B_{\rm{u}}$, respectively.

\begin{figure}[htbp]
\begin{center}
   \includegraphics[width=0.65\textwidth]{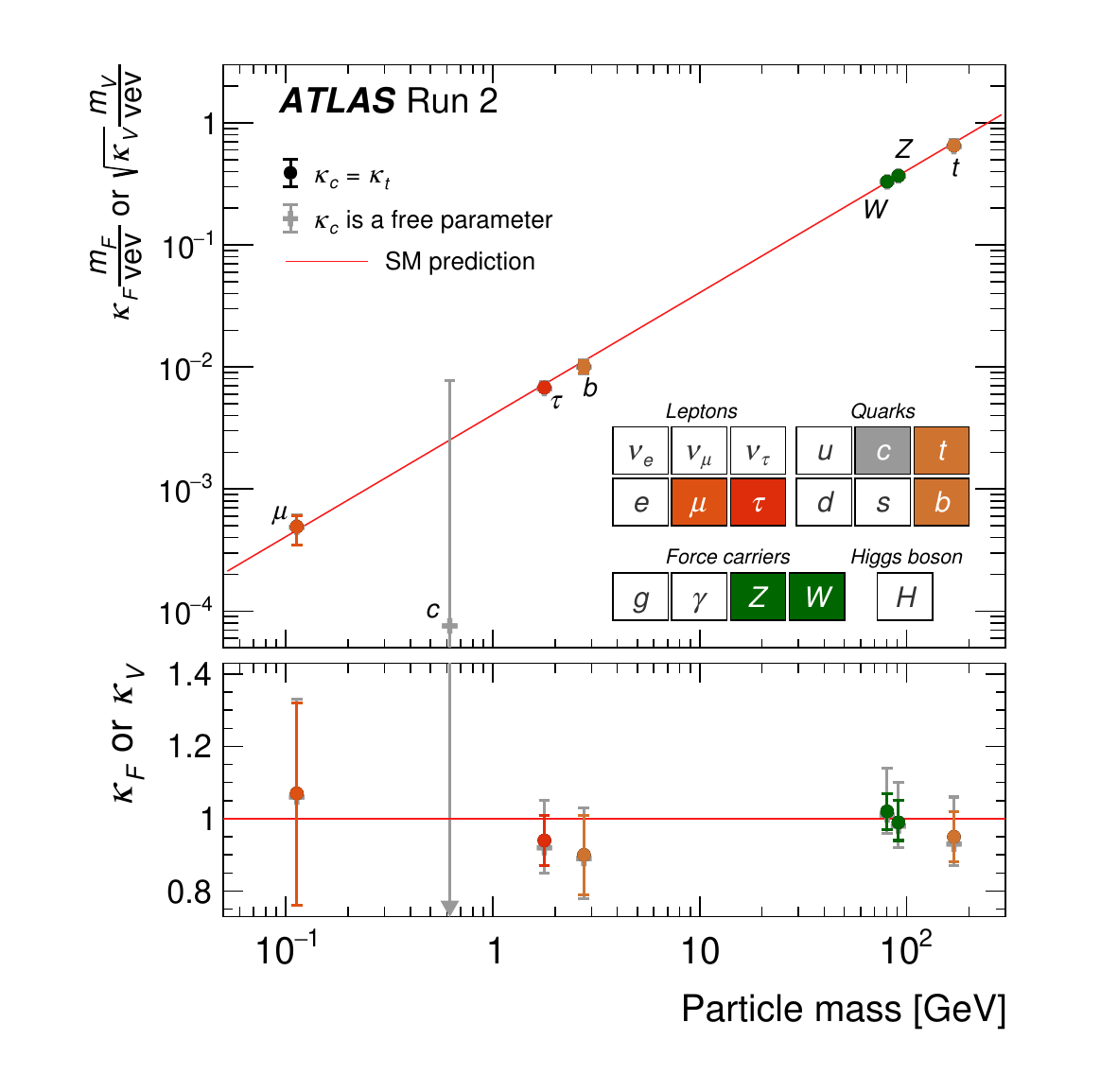}
   \caption{Adapted Higgs boson coupling strength modifiers and their uncertainties (see text) as a function of the boson and fermion masses, as measured by the ATLAS collaboration \cite{ATLAS:2022vkf}. The lower panel shows the values and uncertainties of the modifiers $\kappa_V$ and $\kappa_F$.
   \label{fig:combination_coupling-mass}}
\end{center}
\end{figure}

In an important test the scaling of the couplings of the Higgs boson to the SM particles as a function of their mass is probed and coupling strength modifiers for $W, Z, t, b, c$ and $\tau $ are introduced. It is further assumed that only SM particles contribute to the loop-induced processes and modifications of the fermion and vector boson couplings are propagated through the loop calculations. Invisible and undetected non-SM Higgs boson decays are not considered. In Figure~\ref{fig:combination_coupling-mass} the results obtained by the ATLAS collaboration \cite{ATLAS:2022vkf} are shown in terms of reduced Higgs boson coupling strength modifiers which -- based on the SM coupling relations -- are defined as $\sqrt{ \kappa_V g_V  / 2 v } = \sqrt{ \kappa_V} (m_V / v)$ for weak bosons with mass $m_V$ and 
$\kappa_F g_F  = \kappa_F (m_F / v)$ for fermions with a mass $m_F$, where $g_V$ and $g_F$ are the corresponding absolute coupling strengths and $v$ is the vacuum expectation value of the Higgs field.  Also in this case, the measurements of the coupling strength as a function of mass are in excellent agreement with the SM prediction, which is indicated by the red line. The plot is a clear illustration of a new type of coupling with proportionality to mass as predicted by the implementation of the Higgs mechanism in the SM. 

One can also relax the above-mentioned constraints and allow for the presence of non-SM particles in the loop-induced processes. Such contributions can be parametrised by effective coupling strength modifiers $\kappa_g$, $\kappa_\gamma$ and $\kappa_{Z\gamma}$.
It is also assumed that any potential BSM effects do not modify substantially the kinematic properties of the Higgs boson decay products. Also in this case, the coupling strength modifiers are found to be consistent with 1, indicating no significant deviations from the SM \cite{ATLAS:2022vkf ,  CMS:2022dwd }. When allowing invisible or undetected non-SM Higgs boson decays to contribute to the total width, upper limits of $B_{\rm{inv}} <$0.13 have been set \cite{ATLAS:2022vkf}.

\subsubsection{Higgs boson self-coupling}
As presented in Section \ref{sec:theory predictions}, the measurement of Higgs boson pair production is sensitive to the important Higgs boson self-coupling and thereby to the shape of the Higgs boson potential.  The total production rate for di-Higgs boson production is about three orders of magnitude smaller than single Higgs boson production. The search for Higgs boson pair production is performed in different final states, of which
$b \bar{b} \gamma \gamma $, $b \bar{b} \tau \tau$ and $b \bar{b} b \bar{b}$
are the most sensitive ones. The Higgs boson self-interaction also contributes to other processes via sizeable next-to-leading-order electroweak corrections. It has been shown that the single Higgs boson production cross-sections and branching ratios are also modified if the Higgs boson self-coupling deviates from the SM prediction  \cite{Degrassi:2016wml,Maltoni:2017ims,DiVita:2017eyz,Gorbahn:2016uoy,Bizon:2016wgr,McCullough:2013rea}.

Given the very low production cross-section, the Run-2 dataset is still not sufficient to establish di-Higgs boson production. However, interesting limits have already been obtained by both the ATLAS and CMS collaborations. With the current dataset and combining the searches from all di-Higgs boson channels studied 
the Higgs boson pair production cross-section is found to be less than 2.4 \cite{ATLAS:2022jtk} and 3.4 \cite{CMS:2022dwd} times the SM expectation at the 95\% confidence level by the ATLAS and CMS collaborations, respectively. These limits can be transformed into constraints on the coupling strength modifier $\kappa_\lambda$ for the Higgs boson self-coupling of -0.6 $< \kappa_\lambda <$ 6.6 \cite{ATLAS:2022jtk} and \mbox{-1.2 $< \kappa_\lambda <$ 6.5} \cite{CMS:2022dwd}. This derivation assumes that new physics affects only the Higgs boson self-coupling.
The ATLAS collaboration has also combined information of single Higgs boson cross-section times branching ratio measurements with the cross-section limits from di-Higgs boson production to derive tighter constraints. In this combination, values of $\kappa_\lambda$ are constraint to be in the range \mbox{-0.4 $< \kappa_\lambda < $6.3} with a 95\% confidence level \cite{ATLAS:2022jtk}. Also this derivation assumes that $\kappa_\lambda$ is the only source of physics beyond the SM. If this condition is relaxed, only slightly looser constraints are obtained \cite{ATLAS:2022jtk}.

The important Higgs boson self-coupling parameter remains to be measured. Progress is expected to be made during the upcoming Run 3 (2022 - 2025) at the LHC and at the High-Luminosity LHC (HL-LHC), as discussed in the next section.

\section{Future prospects}

The HL-LHC is scheduled to start operation in 2029. By colliding protons with an instantaneous luminosity of more than three times higher than what is achieved today, the HL-LHC is expected to deliver data corresponding to an integrated luminosity of 3000~\fbs\ by the beginning of the 2040s. Despite the highly challenging experimental environment, such an increased dataset – collected with upgraded detectors – has a huge physics potential: it will give access to rarer physics phenomena and will be critical to achieve high-precision measurements in the Higgs sector as well as for other SM processes. It will also allow to largely extend the sensitivity to sectors of BSM physics that are beyond the present reach, including sensitivity for rarer processes and more complex decay scenarios. 

The study of Higgs boson properties is a central topic in the HL-LHC physics programme. The analyses discussed above will be continued, more differential cross-sections will be measured and possible deviations from the SM can be parametrised in a consistent framework by using an Effective Field Theory (EFT) approach. The main measurements of Higgs boson properties are based on the study of the production and decay modes discussed above. Observation of the rare $H \to \mu \mu$, 
$H \to Z \gamma $ and eventually  $H \to c \bar{c}$ are expected to be reached.  
The production of Higgs boson pairs is a central process to access the Higgs trilinear coupling. Recent studies -- based on the Run-2 experience -- indicate that the ATLAS and CMS experiments can reach a sensitivity to the HH signal of approximately 3$\sigma$ per experiment, leading to a combined sensitivity of about 4$\sigma$ \cite{Cepeda:2019klc} .

While the HL-LHC offers great opportunities,  
some measurements will remain difficult and will leave questions that could be addressed with high precision at future lepton colliders, such as the ILC \cite{Bambade:2019fyw}, CLIC \cite{Brunner:2022usy}, FCC-ee \cite{FCC:2018evy} or CEPC \cite{Gao2021CEPCAS}. On the longer timescale, the discovery reach for direct searches for BSM physics and also the precision in the Higgs sector can be largely extended with a future high-energy hadron collider, such as the FCC-hh \cite{FCC:2018vvp}. 

The trilinear Higgs boson self-coupling is expected to be constrained with an uncertainty of 50\% after the HL-LHC runs, while a combination of FCC-ee and HL-LHC results could reach a precision of about 30\%, and a future hadron collider operating at a centre-of-mass energy of 100~TeV could achieve a clear measurement with a precision of about 5\% \cite{Mangano:2020sao}.

\section{Open questions}
Although the Higgs boson completes the SM of particle physics, it leaves several fundamental questions  unanswered. One significant issue is the fine-tuning or naturalness problem associated with the Higgs sector. This problem arises from the fact that the mass of the Higgs boson assumes a value that is considerably smaller than the magnitude of quantum corrections it would receive in extensions of the SM. This stark contrast between the Higgs boson mass and the expected size of quantum corrections poses a challenge to the theory. If the SM is merely a low-energy approximation of a more fundamental theory, it would be expected for the Higgs mass to be significantly heavier, making its current value appear unnatural.

To illustrate the origin of this problem, the specific example of quantum corrections to the Higgs boson  mass originating from a loop of fermions that interacts with the Higgs boson through a Yukawa interaction is considered: 
\begin{equation}
\mathcal{L} \supset -\lambda_f H \bar{f} f \,.
\end{equation}
Here, $\lambda_f = \sqrt{2} m_f/v$, where $m_f$ represents the mass of the fermion, and $v$ is the vacuum expectation value of the Higgs field.

The one-loop propagator leads to a correction to the Higgs boson mass of the following form (where $N_c$ denotes the colour factor of the fermion):

\begin{equation} 
\Delta m_H^2 = N_c i \lambda_f^2 \int  \frac{d^4 l}{(2\pi)^4} \  \frac{{\rm tr} \left [ ({\slash\!\! l}  + m_t) \left ({\slash \!\! l} + {\slash \!\!\! p} + m_t \right ) \right ]}{(l^2 - m_t^2) \left ((l+p)^2 - m_t^2 \right )}\,.
\end{equation}

This expression diverges quadratically as the loop momentum $|l|$ approaches infinity. The divergent contribution is given by:

\begin{equation}
\Delta m_H^2 = N_c  \frac{i \lambda_f^2}{16 \pi^4} \int  d^4 l \  \frac{4l^2}{l^4} = N_c  \frac{i \lambda_f^2}{4 \pi^4} \left (- i \pi^2 \Lambda^2 \right ) = N_c  \frac{ G_F m_f^2 }{\sqrt{2} \pi^2} \Lambda^2\,.
\end{equation}
Here, $G_F$ represents the Fermi constant, and $\Lambda$ corresponds to the ultraviolet cutoff scale of the theory, which can be interpreted as the energy scale at which new physics effects become significant and modify the behavior of the theory at high energies.

Within the Standard Model (SM), the main contribution to the Higgs boson mass arises from the top quark, but there are also loop effects involving massive bosons that introduce divergent corrections. By considering the most significant one-loop corrections, the following expression is obtained (see e.g.\ Ref.~\cite{Veltman:1980mj}):

\begin{equation}
\frac{\Delta m_H^2}{m_H^2} = \frac{3 G_F}{4 \sqrt{2} \pi^2} \left(\frac{4m_t^2}{m_H^2} - \frac{2m_W^2}{m_H^2} - \frac{m_Z^2}{m_H^2}-1\right)\Lambda^2 \simeq \left( \frac{\Lambda}{500 \  {\rm GeV}}\right)^2\,.
\label{eq:deltamhSM}
\end{equation}

 The hierarchy problem associated with the Higgs boson mass arises from the fact that if the UV cutoff $\Lambda$ is considered to be the Planck mass, the corrections to $m_H$ in Eq.~\eqref{eq:deltamhSM} would be more than 15 orders of magnitude larger than the experimentally measured value of around $125 \, {\rm GeV}$. On the other hand, if there exist new, relatively light particles that act as an effective UV cutoff, they could potentially alleviate the fine-tuning problem in the Higgs sector.
Efforts to resolve or mitigate the fine tuning in the electroweak sector have been a significant focus of theoretical research in particle physics.

For a considerable period of time, this argument has strongly favoured the existence of new physics in the TeV range. However, following the outcomes of Run 1 and Run 2 at the LHC, numerous simple and natural models that aimed to explain the lightness of the Higgs boson are no longer viable or have become less favoured.
An elementary scalar can be naturally light if the mass is protected by a symmetry. The most explored possibility is supersymmetry (SUSY), an internal symmetry which relates fermions to bosons. If SUSY is unbroken, each SM particle has a SUSY partner with the same mass. 
The mass of scalars is then related to the mass of chiral fermions, and hence it is naturally protected from getting quadratically divergent corrections, precisely in the same way as fermion masses in the SM are protected by the chiral symmetry. 
Another possibility is that the Higgs boson is not an elementary, fundamental particle, but is rather a composite state. In this case, the naturalness problem would simply not exist. 

Another question that arises is whether the Higgs sector is as straightforward as postulated in the SM or if additional Higgs bosons exist. Certain models, in contrast to the SM, predict the existence of multiple Higgs bosons and also propose a sharp phase transition during the early cooling stages of the universe. This phase transition is considered one of the necessary conditions for generating the observed baryon asymmetry. The origin of the baryon asymmetry itself represents another fundamental open question in particle physics.

While the SM explains the origin of the masses of all fermions through the Higgs mechanism, it remains unclear why there is such a significant hierarchy in the values of fermion masses. Understanding the underlying origin of this pattern has been a focal point of extensive theoretical efforts over the past few decades. The question at hand is whether an additional structure or pattern exists in a Beyond the Standard Model (BSM) Higgs sector, where this hierarchy emerges more naturally.

The presence and nature of dark matter constitutes another major unsolved mystery. It is currently unknown whether the Higgs boson interacts with dark matter or if the Higgs boson is responsible for the generation of dark matter. Similarly, since the Higgs field is a scalar, it is plausible that the Higgs boson may play a role in driving the process of inflation in the early universe.

One of the intriguing aspects of the Higgs field is whether the Higgs potential is stable, metastable, or unstable. Assuming only the SM, based on three-loop running of the couplings and masses and two-loop matching, one finds no significant tension, at one  $\sigma$ level~\cite{Degrassi:2012ry,Buttazzo:2013uya,Bednyakov:2015sca}, between data and the theoretically predicted critical region. One of the decisive factors is the current uncertainty regarding the top mass, which is likely to be reduced in Run III.

\section{Conclusion}

The discovery of the Higgs boson by the ATLAS and CMS experiments at the LHC has been a groundbreaking milestone in the field of particle physics. The analysis of the complete dataset collected during the first two run periods has led to remarkable advancements in our understanding of the nature of this special particle. Precise measurements of production and decay properties have been carried out and found to be in excellent agreement with the predictions of the SM. Also on the theory side, huge progress has been made in providing precise predictions of production and decays rates, which allow for precision tests.  

However, many measurements, in particular those of rare decays and decays into lighter fermions, are either limited by the available dataset or have not yet been measured. This also applies to the important Higgs boson self coupling which carries important information on the Higgs boson potential.    
Furthermore, there are many open questions in the field, including the naturalness of the Higgs boson mass and its potential connections to other major unresolved questions in fundamental particle theory, such as the presence of dark matter. Addressing these questions is crucial for advancing our understanding of the Higgs boson and its role in the broader context of the universe.

The large dataset which will be collected during the high-luminosity phase at the LHC will allow to make the next step, to increase the measurement precision considerably and thereby to probe smaller deviations from the SM. Beyond this, future high-energy lepton or hadron colliders will play a pivotal role in addressing the open questions and to provide valuable insights into the nature of the Higgs boson and unravel the mysteries that still surround it.

\vskip6pt

\ack{The authors would like to thank the Humboldt Foundation for organising and supporting the inspiring workshop at Kitzb\"uhel which led to interesting interdisciplinary discussions.}

%%%%%%%%%% Insert bibliography here %%%%%%%%%%%%%%

\input{References}
\end{document}

%% file: kjtex.tex
%
%    some useful tex-constructs, to make surviving at Tex level
%    easier....
%    K.J.                                          Aug. 1994
%
%
%    important symbols for real physics (pp collider)
%
%\newcommand{\dzero}{\mbox{D\O}}
%\newcommand{\ppbar}{\mbox{$p\overline{p}$}}
\newcommand{\rts}{\mbox{$\sqrt{s}$}}
\newcommand{\alphas}{\mbox{$\alpha_s$}}
\newcommand{\pt}{\mbox{$p_T$}}
\newcommand{\pT}{\mbox{$p_T$}}
\newcommand{\et}{\mbox{$E_T$}}
\newcommand{\etmiss}{\mbox{$E_T^{miss}$}}
\newcommand{\ptmiss}{\mbox{$P_T^{miss}$}}
\newcommand{\met}{\mbox{\ensuremath{\not \!\! E_T}}}
\newcommand{\pet}{\mbox{\ensuremath{\not \!\! P_T}}}
\newcommand{\ptw}{\mbox{$P_T(W)$}}
\newcommand{\ptz}{\mbox{$P_T(Z)$}}
\newcommand{\deleta}{\mbox{$\Delta \eta \times \Delta \phi$}}
\newcommand{\linteg}{\mbox{$\int{{\cal L} dt}$}}
\newcommand{\abseta}{\mbox{$\mid \eta \mid$}}
\newcommand{\lqcd}{\mbox{$\Lambda_{QCD}$}}
\newcommand{\QSQ}{\mbox{$Q^{2}$}}
\newcommand{\sla}[1]{/\!\!\!#1}
\newcommand{\mevp}{MeV/$c$}
\newcommand{\meve}{MeV/$c^2$}
\newcommand{\gevp}{GeV/$c$}
\newcommand{\geve}{GeV/$c^2$}
\newcommand{\Mcs}{${\rm{MeV/}c^2}$}
\newcommand{\Gcs}{${\rm{GeV/}c^2}$}
\newcommand{\Tcs}{${\rm{TeV/}c^2}$}
\newcommand{\Gc}{${\rm{GeV/}c}$}
\newcommand{\Gcsit}{\ensuremath{\it{GeV/c}^2}}
\newcommand{\Tcsit}{\ensuremath{\it{TeV/c}^2}}
\newcommand{\sw}{\sin\theta_W}
\newcommand{\stbr}{\sigma \cdot BR}
\newcommand{\delphi}{\mbox{$\Delta \phi$}}

%
%   e+ e- physics
%
\newcommand{\epem}{\mbox{${\rm e}^+{\rm e}^-$}}
\newcommand{\epemit}{\mbox{$e^+e^-$}}
\newcommand{\mpmm}{\mbox{$\mu^+\mu^-$}}
\newcommand{\zzero}{\mbox{Z$^0$}}
%
%    B physics
%
%\newcommand{\bs}{\mbox{$B_s^0$}}
\newcommand{\bd}{\mbox{$B_d^0$}}
\newcommand{\bsb}{\mbox{$\bar{B}_s^0$}}
\newcommand{\bdb}{\mbox{$\bar{B}_d^0$}}
\newcommand{\dms}{\mbox{$\Delta m_s$}}
%
%    K physics
%
\newcommand{\ks}{\mbox{$K_S$}}
\newcommand{\kl}{\mbox{$K_L$}}
%
%    general symbols
%
\newcommand{\cl}{\mbox{$95 \% \ CL$}}
\newcommand{\pizero}{\mbox{$\pi^0$}}
\newcommand{\xrad}{\mbox{$X_0$}}
\newcommand{\degr}{\mbox{$^{\circ}$}}
\newcommand{\dsdpt}{\mbox{$frac{d\sigma}{dP_T}$}}
\newcommand{\jpsi}{\mbox{$J/\psi K_s^0$}}
\newcommand{\btojpsi}{\mbox{$B^0_d \rightarrow  J/\psi K_s^0$}}
\newcommand{\btopipi}{\mbox{$B^0_d \rightarrow  \pi \pi$}}
\newcommand{\sbeta}{\mbox{$\sin 2 \beta$}}
\newcommand{\salpha}{\mbox{$\sin 2 \alpha$}}
\newcommand{\dsbeta}{\mbox{$\Delta \sin 2 \beta$}}
\newcommand{\dsalpha}{\mbox{$\Delta \sin 2 \alpha$}}
%
%     Higgs bosons
%
\newcommand{\mH}{\mbox{$m_H$}}
\newcommand{\mZ}{\mbox{$m_Z$}}
\newcommand{\mA}{\mbox{$m_A$}}
\newcommand{\hplus}{\mbox{$H^{\pm}$}}
\newcommand{\mhplus}{\mbox{$m_{H^{\pm}}$}}
\newcommand{\mtautau}{\mbox{$m_{\tau \tau}$}}
%
%     Higgs boson decays
%
\newcommand{\hgg}{\mbox{$H \to  \gamma \gamma$}}
\newcommand{\hbb}{\mbox{$H \to  b \bar{b}$}}
\newcommand{\hcc}{\mbox{$H \to  c \bar{c}$}}
\newcommand{\hzz}{\mbox{$H \to  Z Z$}}
\newcommand{\hzzsfourl}{\mbox{$H \to Z Z^{(*)} \rightarrow 4\ell$}}
\newcommand{\hzzssfourl}{\mbox{$H \to Z Z^{*} \rightarrow 4\ell$}}
\newcommand{\hzzfourl}{\mbox{$H \to Z Z \rightarrow 4\ell$}}
\newcommand{\hww}{\mbox{$H \to  W W $}}
\newcommand{\hwws}{\mbox{$H \to  W W^{(*)}$}}
\newcommand{\hwwss}{\mbox{$H \to  W W^{*}$}}
\newcommand{\hwwsll}{\mbox{$H \to  W W^{(*)} \rightarrow \ell \nu \ell \nu $}}
\newcommand{\hwwssll}{\mbox{$H \to  W W^{*} \rightarrow \ell \nu \ell \nu $}}
\newcommand{\hzzss}{\mbox{$H \to  Z Z^*$}}
\newcommand{\hzzs}{\mbox{$H \to  Z Z^{(*)}$}}
\newcommand{\gamgam}{\mbox{$\gamma \gamma$}}
\newcommand{\htautau}{\mbox{$H \to \tau \tau$}}
\newcommand{\hmumu}{\mbox{$H \to \mu \mu$}}
\newcommand{\hZg}{\mbox{$H \to Z \gamma$}}
\newcommand{\hwwll}{\mbox{$H \to  W W^{*} \rightarrow \ell \nu \ell \nu $}}
%
%   Higgs production
%
\newcommand{\vbf}{\mbox{$qq \rightarrow qq H $}}
%
%     W decays
%
\newcommand{\wlnu}{\mbox{$W \rightarrow \ell \nu $}}
\newcommand{\wenu}{\mbox{$W \rightarrow e \nu $}}
\newcommand{\zll}{\mbox{$Z \rightarrow \ell \ell $}}
\newcommand{\zmumu}{\mbox{$Z \rightarrow \mu \mu $}}
%
%     other LHC processes
%
%
\newcommand{\ttbar}{\mbox{$t \overline{t} $}}
\newcommand{\bbbar}{\mbox{$b \overline{b} $}}
\newcommand{\ccbar}{\mbox{$c \overline{c} $}}
\newcommand{\qqbar}{\mbox{$q \overline{q} $}}
\newcommand{\Zbbbar}{\mbox{$Z b \overline{b} $}}
\newcommand{\ttjets}{\mbox{$t \overline{t} + jets$}}
%
%
%     LHC luminosities
%
\newcommand{\fbs}{\mbox{$\rm{fb}^{-1}$}}
\newcommand{\pbs}{\mbox{$\rm{pb}^{-1}$}}
\newcommand{\fbsit}{\ensuremath{{\it fb}^{-1}}}
\newcommand{\pbsit}{\ensuremath{{\it pb}^{-1}}}

\newcommand{\lhigh}{\mbox{${\cal L} = 10^{34} \ \rm{cm}^{-2} \rm{s}^{-1}$}}
\newcommand{\llow}{\mbox{${\cal L} =  10^{33} \ \rm{cm}^{-2} \rm{s}^{-1}$}}
\newcommand{\lintyear}{\mbox{$\int {\cal L} dt \ = \ 10 \ fb^{-1}$}}
\newcommand{\lintlow}{\mbox{$\int {\cal L} dt \ = \ 30 \ fb^{-1}$}}
\newcommand{\linthigh}{\mbox{$\int {\cal L} dt \ = \  100 \ fb^{-1}$}}
%
%     SUSY parameters
%
%  (i) symbols for particles
%
\newcommand{\chiplus}{\mbox{$\chi^+$}}
\newcommand{\chiminus}{\mbox{$\chi^-$}}
\newcommand{\chizero}{\mbox{$\chi^0$}}
\newcommand{\chipm}{\mbox{$\chi^{\pm}$}}
\newcommand{\chione}{\mbox{$\chi^0_1$}}
\newcommand{\chitwo}{\mbox{$\chi^0_2$}}
\newcommand{\chithree}{\mbox{$\chi^0_3$}}
\newcommand{\chifour}{\mbox{$\chi^0_4$}}
\newcommand{\sel}{\mbox{$\tilde{e}$}}
\newcommand{\smu}{\mbox{$\tilde{\mu}$}}
\newcommand{\stau}{\mbox{$\tilde{\tau}$}}
\newcommand{\slep}{\mbox{$\tilde{\ell}$}}
\newcommand{\slepton}{\mbox{$\tilde{\ell}$}}
\newcommand{\sfer}{\mbox{$\tilde{f}$}}
\newcommand{\sneu}{\mbox{$\tilde{\nu}$}}
\newcommand{\stauone}{\ensuremath{\tilde{\tau}_1 }}
\newcommand{\stautwo}{\ensuremath{\tilde{\tau}_2 }}
\newcommand{\sq}{\mbox{$\tilde{q}$}}
\newcommand{\sgl}{\mbox{$\tilde{g}$}}
\newcommand{\sbot}{\mbox{$\tilde{b}$}}
%
% (ii) SUSY masses
%
\newcommand{\msquark}{\mbox{$m_{\tilde{q}}$}}
\newcommand{\mgluino}{\mbox{$m_{\tilde{g}}$}}
\newcommand{\msneu}{\mbox{$m_{\tilde{\nu}}$}}
\newcommand{\mchiplus}{\mbox{$m_{\tilde{\chi}^+}$}}
\newcommand{\mchi}{\mbox{$m_{\tilde{\chi^0}}$}}
\newcommand{\mzero}{\mbox{$m_0$}}
\newcommand{\mhalf}{\mbox{$m_{1/2}$}}
%
% (iii) SUSY parameters
%
\newcommand{\tanb}{\mbox{$\tan \beta$}}
\newcommand{\matb}{\mbox{$(m_A,\tan \beta )$}}
\newcommand{\matbit}{\ensuremath{{\it (m}_A, {\it tan} \beta {\it )}}}
\newcommand{\tanbit}{\ensuremath{{\it tan} \beta}}
\newcommand{\atau}{\ensuremath{A_{\tau} }}
\newcommand{\phitau}{\ensuremath{\varphi_{\tau} }}
\newcommand{\delm}{\ensuremath{\Delta M}}
\newcommand{\mumtwo}{\ensuremath{(\mu - M_2)}}